 \documentclass[aps,prd,superscriptaddress,showpacs,twocolumn]{revtex4}

\usepackage{amsfonts}
\usepackage{amsmath}
\usepackage{tikz}
\usepackage{graphicx}
\usepackage{xcolor}
\usepackage{amssymb}
\usepackage{soul}
\usepackage{natbib}
\usepackage[normalem]{ulem}
\usepackage{braket}
\usepackage{bm}
\makeatletter
\begingroup
\catcode`\_=\active
\protected\gdef_{\@ifnextchar|\subtextit\subtextup }
\endgroup
\def\subtextit|#1|{\sb{#1}}
\def\subtextup#1{\sb{\mathrm{#1}}}
\AtBeginDocument{\catcode`\_=12 \mathcode`\_=32768}
\makeatother
\renewcommand{\i}{\text i}
\newcommand{\w}{\omega}
\newcommand{\e}{\mathrm{e}}

\newcommand{\eo}{\substack{\mathrm{e} \\[-1.5pt] \mathrm{o}}}
\renewcommand{\Im}{\text{Im}\,}
\renewcommand{\Re}{\text{Re}\,}
\newcommand{\myvec}[1]{\ensuremath{\begin{pmatrix}#1\end{pmatrix}}}

\newcommand{\ten}{\mathbb}
\newcommand{\G}{\ten{G}}

\newcommand{\Gfs}{\mathbb{G}_{vac}^{(0)}}
\newcommand{\nn}{\nonumber \\}

\begin{document}

\title{Purcell effect in chiral environments: A macroscopic QED perspective}

\author{C. S. Rapp}
\affiliation{Institut f\"ur Physik, Universit\"at Kassel, Heinrich-Plett-Stra\ss e 40, 34132 Kassel, Germany}

\author{Janine C. Franz}
\email{janine.franz@u-bordeaux.fr}
\affiliation{Physikalisches Institut, Albert-Ludwigs-Universität Freiburg, Hermann-Herder-Straße 3, 79104 Freiburg, Germany}

\author{Stefan Yoshi Buhmann}
\email{stefan.buhmann@uni-kassel.de}
\affiliation{Institut f\"ur Physik, Universit\"at Kassel, Heinrich-Plett-Stra\ss e 40, 34132 Kassel, Germany}

\author{O. J. Franca}
\email{uk081688@uni-kassel.de}
\affiliation{Institut f\"ur Physik, Universit\"at Kassel, Heinrich-Plett-Stra\ss e 40, 34132 Kassel, Germany}

\begin{abstract}
The Purcell effect describes the modification of the spontaneous decay rate in the presence of electromagnetic media and bodies. In this work, we shed light on the dependencies and magnitude of this effect for chiral materials. Using the framework of macroscopic quantum electrodynamics and Fermi’s golden rule, 
we study a chiral bulk medium with and without local-field corrections, an idealised chiral mirror and a chiral surface. The results imply that the chiral effect is greatest for large transition frequencies, molecules with large optical rotatory strength and media with a strong cross-susceptibility. In the case of a half space, short distances from the molecule to the interface additionally enhance the effect. 

\end{abstract}


\maketitle

\section{Introduction} \label{INTRO}
%


Excited atoms may spontaneously decay and emit their energy in the form of electromagnetic waves \cite{einstein}. In 1946, Purcell discovered through theoretical work that the decay rate of an atom is modified if it is enclosed in a cavity \cite{purcell}. This so-called Purcell effect is dependent on the geometry and material properties of a given medium.

The first experimental verification for the Purcell effect was presented by Drexhage at a conference in 1967 \cite{drex1968}. In a subsequent paper, consistent results for experimental and theoretical decay rates of an atom near a dielectric mirror were published \cite{drex1970}. 

Since then, a variety of geometries for standard media have been analysed theoretically \cite{ddecayrate1,ddecayrate2, ddecayrate3,ddecayrate4} and experimentally \cite{ddecayratex1,ddecayratex2,ddecayratex3}. Recently, the interaction of excited atoms with more complex materials, including magnetoelectric \cite{drlhm}, nonreciprocal \cite{nonreciprocal} and nonlocal \cite{nonlocal} media has been subject of theoretical investigation. In addition, the original setup of Purcell has been generalised to chiral cavities \cite{Yoo-Park}, which paved the way to consider these chiral interactions in the strong-coupling regime. The Purcell effect lies in the weak-coupling regime of light-matter interactions because the coupling between the energy transition of a quantum emitter and an electromagnetic mode is much smaller than all system's decay rates. However, if this coupling is comparable or greater than these dissipative rates, the system enters the strong-coupling regime leading to coherent energy exchange between light and matter \cite{Dovzhenko}. Here the system hosts polaritons and exhibits the Rabi splitting effect, as reported firstly for a single Rydberg atom in a microwave cavity \cite{Rempe}. Such regime is analysed by means of the Jaynes-Cummings model and has been recently considered in Ref.~\cite{Schaefer-Baranov} for chiral interactions. Furthermore, the ultrastrong-coupling regime occurs when the coupling strength is comparable to a significant fraction of the system's frequency and constitutes an active topic too \cite{Dovzhenko,ReviewUSC,Barra-Burillo}, for which the Hopfield Hamiltonian is typically employed. An example of an experimental realization for it was analyzed for transitions in semiconductor heterostructures coupled to metallic and superconducting metasurfaces \cite{Scalari}.
In the following, we will focus on chiral molecules and materials. 

If an object is not superimposable on its mirror image, it is said to be chiral. Object and image are called enantiomers, where we call one \textquotedblleft left-handed\textquotedblright\,and the other \textquotedblleft right-handed\textquotedblright. Different enantiomers are common in nature and the correct handedness of a protein or molecule often determines its functionality. Therefore, the ability to distinguish and separate enantiomers is crucial for various applications in medicine and chemistry. One way to discriminate enantiomers is by their interaction with chiral media, which we introduced before as the Purcell effect. Chiral media are optically active: due to magnetoelectric coupling, the polarization of electromagnetic waves passing through them is rotated. This means that the waves emitted by a decaying chiral molecule will also be altered in the vicinity of a chiral medium, and consequently there will be a chiral interaction between them. Ultimately, we expect the chiral medium to favor the decay of one of the enantiomers.  This reflects a fundamental principle formulated by Curie in 1894: \textquotedblleft The elements of symmetry of the causes must be found in the produced effects\textquotedblright~\cite{curie}.  There exist different types of chiral media, but for this paper we will focus on bi-isotropic media.

Due to its relevance, scattering of light in presence of chiral media has been actively investigated in the last decades, leading to well stablished literature on classical studies \cite{Lakhtakia,Lindell1,Lindell2}. Particularly, the chiral Purcell effect has been previously investigated by means of a semiclassical approach where the electric and magnetic dipoles of a molecule correspond to transitions from some excited state to the ground state (two--level molecule) and are treated as classical sources \cite{Klimov2}. Here the electromagnetic field is a classical solution of the Maxwell equations for different chiral configurations like a single planar interface \cite{Guzatov1}, multi-layer configurations \cite{Guzatov2}, spherical geometry \cite{Klimov1,Hansen,Guzatov3}, a molecule in front of scattering lattices \cite{Hansen}, near two spherical particles \cite{Guzatov5} and even elliptical symmetry \cite{Klimov3} have been analysed. Also nonrelativistic quantum electrodynamics has been applied to study this effect when a chiral molecule near a chiral spherical particle Ref. \cite{Guzatov4}.

In this work, we apply a theory known as macroscopic quantum electrodynamics. In this framework, most quantities of interest can be expressed in terms of the Green’s tensor $\mathbb{G}(\bm r,\bm r',\w)$, which describes the propagation of field--matter excitations of frequency $\w$ between a source point $\bm r'$ and an observation point $\bm r$.

As a small guide to this work: A summary of the necessary theoretical background is presented in Section~\ref{sectheo}. In Section~\ref{secfgr}, a general expression for the decay rate in chiral environments is derived via Fermi's Golden Rule.  In Chapter~\ref{secgeo}, the decay rate for several chiral media under different circumstances is calculated. A summary of the results and an outlook on further investigations can be found in Chapter~\ref{secsum}.

\section{Quantum Electrodynamics in chiral media} \label{sectheo}

The basis of this work is the formalism of macroscopic quantum electrodynamics \cite{SYB1,SYB2}. We use the Maxwell equations and the constitutive relations in chiral media as a starting point
\begin{eqnarray}
\nabla \cdot \underline{\hat{\boldsymbol{D}}}=0 &,& \nabla \times \underline{\hat{\boldsymbol{E}}}-\mathrm{i} \omega \underline{\hat{\boldsymbol{B}}} =\mathbf{0},  \nonumber \\
\nabla \cdot \underline{\hat{\boldsymbol{B}}}=0 &,& \nabla \times \underline{\hat{\boldsymbol{H}}}+\mathrm{i} \omega \underline{\hat{\boldsymbol{D}}} =\mathbf{0}, \label{maxwell}
\end{eqnarray}
where $\hat{\boldsymbol{E}}$ is the electric field, $\hat{\boldsymbol{D}}$ denotes the electric excitation, $\hat{\boldsymbol{B}}$ stands for the magnetic field and $\hat{\boldsymbol{H}}$ labels the magnetic excitation. All of the fields are given in the frequency realm, expressed by the underline. A field $\underline{\hat{\boldsymbol{F}}}$ can be transformed back into spatial coordinates via $\hat{\boldsymbol{F}} = \int_0^\infty\text d\w \underline{\hat{\boldsymbol{F}}} \text e^{-i\w t} + \text{H.c.}$

The chiral constitutive relations for the fields $\underline{\hat{\bm D}}$ and $\underline{\hat{\bm B}}$ are given as shown \cite{knoise,pchgreent}: 
\begin{equation}
\begin{aligned}
\underline{\hat{\boldsymbol{D}}}&=\varepsilon_0 \varepsilon \underline{\hat{\boldsymbol{E}}}+\underline{\hat{\boldsymbol{P}}}_{\mathrm{N}}+\mathrm{i} \frac{\kappa}{c}\left(\underline{\hat{\boldsymbol{H}}}+\underline{\hat{\boldsymbol{M}}}_{\mathrm{N}}\right) \;,\\
\underline{\hat{\boldsymbol{B}}}&=\mu_0 \mu\left(\underline{\hat{\boldsymbol{H}}}+\underline{\hat{\boldsymbol{M}}}_{\mathrm{N}}\right)-\mathrm{i}\frac{\kappa}{c} \underline{\hat{\boldsymbol{E}}} \;, \label{constrel}
\end{aligned}
\end{equation}
where the electric permittivity $\varepsilon$ and magnetic permeability $\mu$ describe the electromagnetic properties of a material. To describe the chiral behavior, we introduce the chiral cross-susceptibility $\kappa$, which we will assume to be a real variable unless otherwise specified \cite{emchirality}. $\underline{\hat{\bm P}}_N$ and $\underline{\hat{\bm M}}_N$ stand for the noise polarization and magnetization respectively, generated by quantum fluctuations. They can be written in terms of fundamental bosonic operators. In the literature, the constitutive relationships (\ref{constrel}) are identified by the name Boys--Post \cite{Lakhtakia,Lindell1}, which are more fundamental from the point of view of field quantization \cite{Guzatov4,Horsley}. Nevertheless, the Drude--Born--Fedorov constitutive relationships \cite{Lakhtakia,Lindell1} are another set of them and are frequently employed in classical and semiclassical studies of chiral media \cite{Guzatov1,Guzatov2,Guzatov3,Hansen,Bohren1,Bohren2,Klimov1,Klimov2,Klimov3}. 

By combining the Maxwell equations (\ref{maxwell}) and constitutive relations (\ref{constrel}), we find the inhomogeneous  Helmholtz equation for the electric field:
\begin{align}
\left[\frac{1}{\mu}\nabla\times\nabla\times-2\frac{\w}{c}\frac{\kappa}{\mu}\nabla\times-\frac{\w^2}{c^2}\left(\varepsilon-\frac{\kappa^2}{\mu}\right)\right]\underline{\hat{\bm E}}=\i \mu_0\w\underline{\hat{\bm j}}_N.
\label{helmholtz}
\end{align}

The Green's tensor $\mathbb{G}(\bm r, \bm r',\w)$ is defined as the solution to this equation:
\begin{equation}
\begin{aligned}
\bigg[\frac{1}{\mu}\nabla\times\nabla\times-2\frac{\w}{c}\frac{\kappa}{\mu}\nabla&\times-\frac{\w^2}{c^2}\left(\varepsilon-\frac{\kappa^2}{\mu}\right)\bigg] \mathbb{G}(\bm r, \bm r',\w)\\
=& \mathbb{I} \delta(\bm r-\bm r'). \label{helmholtzsol}
\end{aligned}
\end{equation}

Although these definitions of $\varepsilon$, $\mu$ and $\kappa$ are most convenient for our purposes, the equivalent definitions $\varepsilon'=\varepsilon-\kappa^2/\mu$ and $\kappa'=\kappa/\mu$ are widespread in literature, e.g. in the work \cite{chbulk}.

With this, the fields $\underline{\hat{\boldsymbol{E}}}$ and $\underline{\hat{\boldsymbol{B}}}$ in question can be expressed through the Green's tensor and the bosonic creation $\underline{\hat{\boldsymbol{f}}}^\dagger_\lambda$ and annihilation operators $\underline{\hat{\boldsymbol{f}}}_\lambda$. The electric field in spatial coordinates is given by
\begin{equation}
\begin{aligned}
\hat{\bm E}(\bm r)=&\int_0^{\infty} \mathrm{d} \omega \sum_{\lambda=e, m} \int \mathrm{d}^3 r^{\prime} \mathbb{G}_\lambda\left(\boldsymbol{r}, \boldsymbol{r}^{\prime}, \omega\right) \cdot \underline{\hat{\boldsymbol{f}}}_\lambda\left(\boldsymbol{r}^{\prime}, \omega\right)
\\&+\text { H.c., } \label{efg}
\end{aligned}
\end{equation}
hence the magnetic field can be calculated via Faraday's law
\begin{equation}
\begin{aligned}
\hat{\boldsymbol{B}}(\boldsymbol{r})=&\int_0^{\infty} \frac{\mathrm{d} \omega}{\mathrm{i} \omega} \sum_{\lambda=e, m} \int \mathrm{d}^3 r^{\prime} \nabla \times \mathbb{G}_\lambda\left(\boldsymbol{r}, \boldsymbol{r}^{\prime}, \omega\right) \cdot \underline{\hat{\boldsymbol{f}}}_\lambda\left(\boldsymbol{r}^{\prime}, \omega\right) \\
&+\text { H.c.}\,,\label{mfg}
\end{aligned}
\end{equation}
where $\lambda$ denotes the modes and we defined \cite{SYB1}
\begin{eqnarray}
\G_e \left(\boldsymbol{r}, \boldsymbol{r}^{\prime}, \w \right) &=& \i \frac{ \w^2 }{ c^2 } \sqrt{ \frac{ \hslash }{ \pi \varepsilon_0 } \text{Im}\,\varepsilon \left(\boldsymbol{r}^{\prime}, \w \right) }\, \G \left(\boldsymbol{r}, \boldsymbol{r}^{\prime}, \w \right) \,, \\
\G_m \left(\boldsymbol{r}, \boldsymbol{r}^{\prime}, \w \right) &=& -\i \frac{ \w^2 }{ c^2 } \sqrt{ \frac{ \hslash \text{Im}\,\mu \left(\boldsymbol{r}^{\prime}, \w \right) }{ \pi \varepsilon_0 |\mu\left(\boldsymbol{r}^{\prime}, \w \right)|^2 } } \G \left(\boldsymbol{r}, \boldsymbol{r}^{\prime}, \w \right) \times \overleftarrow{\nabla} \,. \nonumber\\
\end{eqnarray}
The operation $\times \overleftarrow{\nabla}$ means that the derivative acts on the second spatial variable where the source is situated, e.g. $[{\bf\mathbb{T}} \times \overleftarrow{\nabla}]_{i j}\left(\mathbf{r}, \mathbf{r}^{\prime}\right)=\varepsilon_{j k l} \partial_l^{\,\prime} T_{i k}\left(\mathbf{r}, \mathbf{r}^{\prime}\right)$.

Additionally, the coefficients obey the following integral relation that will be useful for us~\cite{SYB1}:
\begin{equation}
\begin{aligned}
\sum_{\lambda=e, m} \int \mathrm{d}^3 s\, & \mathbb{G}_\lambda(\boldsymbol{r}, \boldsymbol{s}, \omega) \cdot \mathbb{G}_\lambda^{*\,\mathrm{T}}\left(\boldsymbol{r}^{\prime}, \boldsymbol{s}, \omega\right)  \\
=&\frac{\hslash \mu_0}{\pi} \omega^2 \operatorname{Im} \mathbb{G}\left(\boldsymbol{r}, \boldsymbol{r}^{\prime}, \omega\right).\label{intrel}
\end{aligned}
\end{equation}


\section{Decay rate} \label{secfgr}

Fermi's golden rule can be derived via time-dependent perturbation theory \cite{sakurai}. It states that the decay rate of an excited system is given by the sum of all transitions to lower levels:
\begin{equation}
\Gamma=\frac{2\pi}{\hslash}\sum_KV_{IK} \cdot V_{KI}\delta(E_{IK}). \label{chd}
\end{equation}
In this expression $V_{IK}$ is a matrix element of the interaction potential and $E_{IK}=\hslash\w_{IK}$ with $\w_{IK}$ standing for the transition frequencies. For our system of an excited chiral molecule in a chiral environment, we have the interaction potential $\hat V=-\hat{\bm {d}}\cdot \hat{\bm E}(\bm r_M)-\hat{\bm {m}}\cdot \hat{\bm B}(\bm r_M)$ that has both electric and magnetic dipoles, reflecting the chiral nature of a molecule. We define a basis of Fock states: 
\begin{equation}
\begin{aligned}
\ket{I} =& \ket{e, \{0\}} = \ket{e}\ket{\{0\}} \\
\ket{K} =& 	\ket{g, \bm 1_\lambda(\bm r, \w)} = \ket{g}\ket{\bm 1_\lambda(\bm r, \w)}.
\end{aligned}
\end{equation}
In the initial state $\ket{I}$, the molecule is in the excited state $\ket{e}$, while the field is in the ground state $\ket{\{0\}}$. In the final state $\ket{K}$, these roles have been reversed. The dipole transition matrix elements are given by
\begin{equation}
\bm d_{IK} = \braket{e|\hat{\bm{d}}|g}, \qquad
\bm m_{IK} = \braket{e|\hat{\bm{m}}|g}.
\end{equation}
At this point we calculate the matrix elements of $\hat V$, employing the expansions (\ref{efg}) and (\ref{mfg}) electric and magnetic fields. We hereby obtain a purely electric term responsible for the usual electric contribution. Then there is a purely magnetic term, which we will neglect due to the lower magnitude of the magnetic transition dipole compared with the electric transition dipole. Finally we have cross-terms, which are responsible for a chiral response. These have the following structure:
\begin{equation}
\begin{aligned}
&\braket{I |\hat{\bm {m}}\cdot \hat{\bm B}(\bm r_M)|\text K} \braket{\text K|\hat {\bm {d}}\cdot \hat{\bm E}(\bm r_M)| I} =\\
\frac{1}{\i \w}\bm m_{IK}\cdot&\sum_{\lambda}\nabla\times\mathbb{G}_{\lambda}(\bm r_M,\bm r,\w)\cdot \mathbb{G}_{\lambda}^{*\,\mathrm{T}}(\bm r_M,\bm r,\w) \cdot\bm d_{IK}^*
\end{aligned}
\end{equation}
and
\begin{equation}
\begin{aligned}
&  \braket{I |\hat {\bm {d}}\cdot \hat{\bm E}(\bm r_M)|\text K}\braket{\text K|\hat {\bm {m}}\cdot \hat{\bm B}(\bm r_M)| I} =\\
\frac{1}{\i \w}\bm d_{IK}\cdot&\sum_{\lambda}\mathbb{G}_{\lambda}(\bm r_M,\bm r,\w)\cdot \mathbb{G}_{\lambda}^{*\,\mathrm{T}}(\bm r_M,\bm r,\w) \times\overleftarrow\nabla\cdot\bm m_{IK}^*.
\end{aligned}  
\end{equation}
By simplifying, using the integral relation (\ref{intrel}) and substituting into the decay rate (\ref{chd}), we obtain the chiral contribution
\begin{equation}
\begin{aligned}
\Gamma_{ch}=&\frac{2\mu_0\w_{IK}}{\i\hslash} \left[\vphantom{\overleftarrow\nabla}\bm m_{IK}\cdot\nabla\times\text{Im }\mathbb{G}(\bm r_M,\bm r_M,\w_{IK})\cdot\bm d_{IK}^*\right.\\
&\left.+\bm d_{IK}\cdot\text{Im }\mathbb{G}(\bm r_M,\bm r_M,\w_{IK})\times\overleftarrow\nabla\cdot\bm m_{IK}^*\right]. \label{chdr0}
\end{aligned}
\end{equation}

By exploiting the Onsager reciprocity of the Green's tensor valid for reciprocal media $\mathbb{G}(\mathbf{r},\mathbf{r}',\omega) = \mathbb{G}^{\mathrm{T}}(\mathbf{r}',\mathbf{r},\omega)$ and some algebraic manipulation, we discover:
\begin{equation}
\begin{aligned}
\bm d_{IK} \cdot\text{Im }\mathbb{G}\times\overleftarrow\nabla\cdot\bm m_{IK}^* =-(\bm m_{IK}\cdot \nabla\times\Im\mathbb{G}\cdot \bm d_{IK}^*)^*.\label{imoderso}
\end{aligned}
\end{equation}
Thus, we find the more compact expression
\begin{equation}
\Gamma_{ch}=\frac{4\mu_0\w_{IK}}{\hslash }\text{Im}\left[\bm m_{IK}\cdot \nabla\times\text{Im }\mathbb{G}(\bm r_M,\bm r_M,\w_{IK})\cdot \bm d_{IK}^*\right]. \label{chdr}
\end{equation}

If either the environment is not chiral, $\nabla\times\text{Im }\mathbb{G}=\boldsymbol{0}$, or if the object is achiral, i.e. does not exhibit an electric or magnetic dipole simultaneously, there will not be a chiral component to the decay rate. Additionally, we observe that $\bm m_{IK}\cdot \nabla\times\text{Im }\mathbb{G}\cdot \bm d_{IK}^*$ must be complex, meaning that polarizations will play a key role to give a nonzero result. Our macroscopic-QED result agrees with that obtainable from a semiclassical calculation with arbitrary multipole contributions when focussing on mixed electric/magnetic dipole contributions \cite{Klimov2}.

The total decay rate is the sum of the electric and chiral components
\begin{equation}\label{whole dr}
\Gamma=\Gamma_{el}+\Gamma_{ch},
\end{equation}
where the electric contribution is known \cite{SYB2}:
\begin{equation}
\Gamma_{el}=\frac{2\mu_0}{\hslash}\omega_{IK}^2\bm d_{IK}\cdot\operatorname{Im} \mathbb{G}\left(\boldsymbol{r}_M, \boldsymbol{r}_M, \omega_{IK}\right) \cdot\bm d_{IK}^*.\label{ddr}
\end{equation}
A comparison between the two contributions immediately shows the necessity of a chiral object to be able to distinguish between the chiral states of another object, i.e. the Curie principle is manifestly fulfilled \cite{curie}. 



\section{Decay rates for chiral environments}\label{secgeo}

In this section, we will calculate the explicit decay rate for a few simple chiral environments. We will start with a chiral bulk medium, then we will analyze a perfect chiral mirror and finally we will have a look at a chiral surface with less idealised properties.\\

\subsection{Chiral bulk medium}\label{secchiralbulk}
The chiral bulk medium is an infinitely extended homogeneous medium with chiral properties. We denote these chiral properties with a chiral cross-susceptibility $\kappa(\w)$, which gives rise to the set of chiral constitutive relations (\ref{constrel}). There are different conventions for the sign and symbol of the chiral cross-susceptibility; one may encounter it as $\chi$ or even having the opposing sign. Using $\kappa>0$ for a right-handed and $\kappa<0$ for a left-handed medium with our definitions, wave numbers corresponding to circular polarized modes in the chiral medium can be written as \cite{chbulk}:
\begin{equation}
\begin{aligned}
&k_{-}=k_0 \left(n_r+\kappa\right) \\
&k_{+}=k_0 \left(n_r-\kappa\right),\label{cwv}
\end{aligned}
\end{equation}
where $k_0=\w/c$ is the field's vacuum wave number, $n_r=\sqrt{\varepsilon\mu}$  is the non-chiral refractive index and the subscripts +,- indicate left- or right-circularly polarized waves.  

In Ref.~\cite{chbulk}, the Green's tensor for a chiral bulk medium is derived in terms of spherical vector wave functions $\bm W_{\mathit{mn}}$   and $\bm V_{\mathit{mn}}$, which for their radial part are expressed in terms of spherical Hankel functions. The full expressions for $\bm W_{\mathit{mn}}$ and $\bm V_{\mathit{mn}}$ can be found in Appendix \ref{A}.  We may choose the source point $\bm r'$ at the origin, which considerably simplifies the functions. We observe that $\bm V_{\mathit{mn}}$ and $\bm W_{\mathit{mn}}$ with $n>1$ vanish in this case. This results in
\begin{equation}
\begin{aligned}
\mathbb{G}^{(0)}(\bm r,\bm r',\omega) = & \frac{3\i\mu}{\pi(k_++k_-)} \sum_{\mathit{m}=0}^1 \left[ k_+^2\bm V_{\eo \mathit{m}1}^{(1)}(k_+) \bm V_{\eo \mathit{m}1}^{(0)}(k_+) \right. \\
&\left. +k_-^2 \bm W^{(1)}_{\eo \mathit{m}1}(k_-) \bm W_{\eo \mathit{m}1}^{(0)}(k_-) \right] . \label{chbulk1}
\end{aligned}
\end{equation}
%

 The functions $\bm V, \bm W$ satisfy the following relation for the curl:
\begin{equation}
\begin{aligned}
\nabla\times \bm V(k_+)&=k_+\bm V(k_+)\;, \\
\nabla\times \bm W(k_-)&=-k_- \bm W(k_-)\;.
\end{aligned}\label{krunter}
\end{equation}

With this relation, the curl of the Green's tensor can easily be identified as
\begin{equation}
\begin{aligned}
&\nabla \times \mathbb{G}^{(0)}(\bm r, \bm r', \omega) =\frac{3\i\mu k_+^3 }{\pi(k_++k_-)} \\
&\times\sum_{\mathit{m}=0}^1 [ \bm V_{\eo \mathit{m}1}^{(1)}(k_+) \bm V_{\eo \mathit{m}1}^{(0)}(k_+) - \frac{ k_-^3 }{ k_+^3 } \bm W^{(1)}_{\eo \mathit{m}1}(k_-) \bm W_{\eo \mathit{m}1}^{(0)}(k_-) ] .
\end{aligned}
\end{equation}

To take the imaginary part and present it in a tangible format, we substitute the functions $\bm V$ and $\bm W$. In spherical coordinates, we then obtain a matrix with the following form:

\begin{multline}
\nabla \times \mathbb{G}^{(0)}(\bm r, \bm r', \w)
\\
=\sum\limits_{P=\pm} 
\frac{\mu(\w)\e^{\i {k_P} \rho} (1-\i {k_P} \rho)  }{2 \pi  \rho^3 ({k_+}+{k_-})}
\\
\times 
\left[
	\sigma_{P}  
\mathbb{I} 
+
\frac{ k_P \rho}{2}
( \mathbf{e}_\theta \otimes\mathbf{e}_\phi-
\mathbf{e}_\phi\otimes \mathbf{e}_\theta)
\right] \,,
\end{multline}
where $\rho=|\bm r - \bm r' |$ and $\sigma_{P} = \pm 1$ for $P = +$ and $P = -$, respectively. To calculate the imaginary part,  we assume a non-absorbing medium with $\varepsilon$, $\mu$ and $\kappa$ real, we take the coincidence limit with $\rho\rightarrow0$ and approximate the result with a Taylor series: 
\begin{equation}\label{rotG_bulk}
\begin{aligned}
&\nabla \times \Im\mathbb{G}^{(0)}(\bm r, \bm r, \w)=\frac{\mu(\w)(k_+^3-k_-^3)}{6(k_++k_-)\pi} \mathbb{I} \;.
\end{aligned}
\end{equation}

We then substitute the result for the Green's tensor back into the chiral contribution (\ref{chdr}). In lowest order of $\kappa$, we obtain
\begin{equation}
\begin{aligned}
\Gamma_{ch}^{(0)}&=-\frac{2\mu n_r\w_{IK}^3\kappa}{\hslash\varepsilon_0\pi c^4}\Im(\bm d_{IK}\cdot \bm m_{IK}^*)\;.
\end{aligned}
\end{equation}

Written in terms of the optical rotatory strength $R_{IK}=\Im (\bm d_{IK}\cdot\bm m_{IK}^*)$ \cite{optrot} the chiral contribution reads
\begin{equation}
\Gamma_{ch}^{(0)}=-\frac{2\mu n_r\w_{IK}^3\kappa R_{IK}}{\hslash\varepsilon_0\pi c^4}\;.\label{chdrbulk}
\end{equation}

Some comments regarding these results are now in order. Due to the vectorial structure of $R_{IK}$, the electric and magnetic dipoles will contribute maximally when they are colinear. For achiral molecules, $R_{IK}$ has the value 0 and the decay rate vanishes. The expression (\ref{chdrbulk}) changes in sign for different enantiomers. The sign of the chiral decay rate depends on relative optical rotation the molecule and medium. To establish a correspondence between the respective parameters $R_{IK}$ and $\kappa$, we compare the constitutive relations (\ref{constrel}) of the chiral medium with the magneto-electric response 
\begin{equation}
\begin{aligned}
\underline{\boldsymbol{d}} & = \chi_{ee}\underline{\boldsymbol{E}} + \chi_{em} \underline{\boldsymbol{B}}/c \;, \\
\underline{\boldsymbol{m}}/c &= \chi_{me}\underline{\boldsymbol{E}} + \chi_{mm} \underline{\boldsymbol{B}}/c 
\end{aligned}
\end{equation}
of a single molecule, with $\underline{\boldsymbol{d}}$ and $\underline{\boldsymbol{m}}$ denoting its electric and magnetic dipole moments, respectively. The comparison reveals that $\kappa$ corresponds to \cite{Marachevsky2}
\begin{equation}
\i \chi_{em} = - \frac{ 2 }{ 3\hslash } \sum_K \frac{ R_{IK} \w }{ \w_K^2 - \w^2} \;. 
\end{equation}
We conclude that the chiral decay rate is positive if molecule and medium are both dextrorotatory or both levorotatory (i.e.~either $\kappa>0$ and $R_{IK}<0$ or $\kappa<0$ and $R_{IK}>0$) and negative if they exhibit opposite optical rotation (either $\kappa>0$ and $R_{IK}>0$ or $\kappa<0$ and $R_{IK}<0$). 

The electric contribution to the decay rate for a comparable dielectric bulk medium is given by \cite{SYB2}
\begin{equation}
\Gamma_{el}^{(0)}=\frac{\mu(\w_{IK}) n_r\w_{IK}^3|\bm d_{IK}|^2}{3\pi\varepsilon_0\hslash c^3} \label{ddrbulk}
\end{equation}
 and is insensitive to changes in handedness, i.e.~it is achiral. A quantity that may be taken notice of is the degree of discrimination $S$ \cite{degdisc} between enantiomers with positive ($\Gamma_+$) and negative ($\Gamma_-$) rotatory strength:
\begin{equation}
S=\frac{\Gamma_+-\Gamma_-}{\Gamma_++\Gamma_-}=\frac{\Gamma_{disc}}{\Gamma_{nd}}\in \left[-1,1\right] \;,
\label{def_s}
\end{equation}
where $\Gamma_{disc}$ discriminates between enantiomers and $\Gamma_{nd}$ is the non-discriminating electric decay rate. This degree of discrimination $S$ resembles the ratio of spontaneous emission rate from a chiral molecule to the spontaneous emission rate from the same chiral molecule in vacuum reported in Refs.~\cite{Guzatov1,Guzatov2,Guzatov3,Hansen,Klimov2,Klimov3}. Due to the fact that different chirality only produces a sign change for the chiral contribution, we deduce that $S$ is nothing other than the ratio of the purely chiral contribution and the electric contribution to the decay rate: 
\begin{equation}\label{squo}
\begin{aligned}
S^{(0)} = \frac{\Gamma_{ch}^{(0)} }{ \Gamma_{el}^{(0)} } = -\frac{6\kappa R_{IK}}{c|\bm d_{IK}|^2}\;,
\end{aligned}
\end{equation}
whose magnitude is modulated by $\kappa$ and the ratio of the optical rotatory strength to the squared magnitude of the electric dipole. Using the order-of-magnitude estimate of Ref.~\cite{md}, it is about 6 times the fine structure constant $\alpha$ for a typical molecule.  \\



\subsubsection{Local-field corrections}
Previously, the molecule was suspended in the medium without regard for the space it occupies. In practice, the nearest particles around the molecule are at least the length of a particle away. To take this into account we use a model where the molecule is placed in a vacuum sphere inside the medium, which is known as the Onsager real cavity model \cite{degdisc,pbubble,Fiedler,JanineNJP,Onsager}. Here we would like to remark the distinction between our work and previous ones devoted to the spherical geometry. Although the classical scattering of electromagnetic waves by chiral spheres has been studied on Ref.~\cite{Lindell2} providing the basis of the Purcell effect by means of the semiclassical approach of Refs.~\cite{Guzatov3,Klimov1} and the quantum treatment of Ref.~\cite{Guzatov4}, all these studies as well as Ref.~\cite{Hansen} always consider the chiral source to be outside the chiral sphere, which enables us to analytically and numerically study the decay rate of spontaneous emission of the chiral molecule. Meanwhile, in our case the chiral source is surrounded by the chiral medium as discussed above, which allows us to examine the local-field corrections in this section.

The contributions for the decay rate are then twofold: The vacuum decay rate and the scattering part that arises from the interaction of the molecule with the sphere surface. The derivation for the Green's tensor for this scenario is given in Appendix~\ref{B}. The result is in agreement with the recently obtained chiral local-field corrections for energy transfer derived in Ref.~\cite{thesis_janine}. 

In the case of real parameters, there is only a single contribution to the Green's tensor, cf. Eq.~(\ref{curl_g_lfc}):
\begin{equation}
\begin{aligned}
\nabla \times \Im\mathbb G^{\text{lfc}} &= \frac{\kappa \w^2}{\pi c^2}f_0\mathbb{I} \;, \\
&= f_0 \nabla \times \Im\mathbb{G}^{(0)}(\bm r, \bm r, \w) \; , \label{rotG_lfc_real}
\end{aligned}
\end{equation} 
where $f_0$ is defined in Eq.~(\ref{f0App}) and can be rewritten as follows
\begin{multline}
f_0(\varepsilon,\mu,\kappa) = \frac{ \left( \mathcal{F}_{\varepsilon\mu}+\kappa\mathcal{F}_{\kappa} \right)k_-^3 -  \left( \mathcal{F}_{\varepsilon\mu}-\kappa\mathcal{F}_{\kappa} \right)k_+^3 }{ k_-^3-k_+^3 } , \label{f0}
\end{multline}
with
\begin{equation}\label{curlyFs}
\begin{aligned}
\mathcal{F}_{\varepsilon\mu} &= \left(\frac{3\varepsilon}{2\varepsilon+1}\right)\left(\frac{3}{2\mu+1}\right) \;,\\
\mathcal{F}_{\kappa} &= 9\sqrt{ \frac{\varepsilon}{\mu} } \frac{ 4n_r^2 -1 }{ (2\varepsilon+1)^2(2\mu+1)^2 } \; .
\end{aligned}
\end{equation} 

Equation (\ref{rotG_lfc_real}) correctly reproduces the uncorrected case (\ref{rotG_bulk}) by setting $f_0=1$ in Eq.~(\ref{f0}) or equivalently $\mathcal{F}_{\epsilon\mu}=1$ and $\mathcal{F}_{\kappa}=0$ in Eqs.~(\ref{curlyFs}). Within leading zero-order in $\kappa$, $f_0$ reduces to $f_0\simeq f_0(\varepsilon,\mu)=\mathcal{F}_{\varepsilon\mu}$. \\

Thus, the chiral contribution taking into account local-field corrections reads
\begin{equation}\label{bulklfc}
\Gamma_{ch}^{\text{lfc}} = f_0(\varepsilon,\mu,\kappa) \Gamma_{ch}^{(0)} \;,
\end{equation}
where $\Gamma_{ch}^{(0)}$ is given in Eq.~(\ref{chdrbulk}). As can be identified swiftly, taking into account local-field corrections yields a factor for this contribution that to leading order in $\kappa$ is dependent only on the permittivity $\varepsilon$ and permeability $\mu$ of the medium, while not influencing the overall structure of the interaction between the medium and the molecule.

%
%
%
%


%


For further understanding, we compare with the decay rate containing local-field corrections for a dielectric bulk. It  can be found in the article \cite{pbubble}, obtained also via Onsager's real cavity model, and it reads
\begin{equation}
\Gamma_{el}^{\text{lfc}} = \left(\frac{ 3\varepsilon }{ 2\varepsilon +1 }\right)^2  \Gamma_{el}^{(0)} \;,
\end{equation}
where $\Gamma_{el}^{(0)}$ was introduced in Eq.~(\ref{ddrbulk}).

The corresponding degree of discrimination~(27) is then given by
\begin{equation}\label{slfc}
S^{\text{lfc}} = \left(\frac{ 3\varepsilon }{ 2\varepsilon +1 }\right)^{-2} f_0\, S^{(0)}\;,
\end{equation}
where we expressed this result in terms of the degree of discrimination without any local-field corrections (\ref{squo}), to make more evident that their differences only lie in a factor that depends on the medium properties $\varepsilon, \mu$ and $\kappa$. From Eq.~(\ref{slfc}), the degree of discrimination without local-field corrections is recovered by setting the factor $f_0(\varepsilon,\mu,\kappa)$ to 1.

\subsubsection{Local-field corrections in an absorbing medium}

Lastly in the context of a bulk medium, we have a look at the decay rate for a medium with complex-valued $\varepsilon ,\mu$ and $\kappa$ --- corresponding to absorbing media. 
%
 In the limit of small sphere radius $a \ll 1$, we find that the imaginary part of the curl of the Green's tensor is given by
\begin{multline}
\nabla\times\Im\mathbb{G}^{\text{lfc}} 
= \Im \left[ 
\frac{3 \kappa }{2 \pi  a^3 k_0 (2 \mu +1) (2 \varepsilon +1)}
\right] \ten{I} 
\\ + \mathcal{O}(a^{-1}) + \mathcal{O}(1)
\,, \label{curlG_abs}
\end{multline}
see Appendix~\ref{B} for its derivation.
The dominating term $\propto 1/a^3$ only contributes to the imaginary part for complex medium parameters. The same holds true for the term of $\mathcal{O}(a^{-1})$. If we were to use real parameters here, only the $\mathcal{O}(1)$-term already given in Eq.~\eqref{rotG_lfc_real} would contribute.

Substituting Eq.~(\ref{curlG_abs}) into Eq.~(\ref{chdr}), the chiral contribution to the decay rate for a molecule in a chiral absorbing bulk medium with local-field corrections yields:
\begin{eqnarray}
&&\Gamma_{ch}^{\text{lfc}} = \frac{6 R_{IK} }{a^3 \pi \hslash \varepsilon_0 c |2\varepsilon+1|^2 |2\mu+1|^2  } \nonumber\\
&& \times \left[ 2\Re\kappa \left( 2\Re\varepsilon +1\right) \Im \mu + 2\Re\kappa \Im \varepsilon \left( 2\Re\mu +1\right) \right.\nonumber\\
&& \left. -\Im\kappa \left( 2\Re\varepsilon +1\right) \left( 2\Re\mu +1\right) + 4 \Im\kappa \Im \varepsilon \Im \mu \right] . \qquad
\end{eqnarray}

However, most materials present $\mu=1$, which immediately leads us to the simplification;
\begin{eqnarray}
\Gamma_{ch}^{\text{lfc}} &=& \frac{2 R_{IK} }{a^3 \pi \hslash \varepsilon_0 c |2\varepsilon+1|^2 } \nonumber\\
&& \times \left[ 2\Re\kappa \Im \varepsilon - \Im\kappa\left( 2\Re\varepsilon +1\right) \right]\;.\label{dr_abs_lfc}
\end{eqnarray}

Now, we proceed to evaluate the discrimination between the enantiomers. To allow us to calculate the degree of discrimination $S$, we require the analogue contribution for an absorbing dielectric medium, given by \cite{pbubble} 

\begin{equation}
\begin{aligned}
\Gamma_{el}^{\text{lfc}} = \frac{ |\bm d_{IK}|^2 }{ \pi\hslash\varepsilon_0 a^3 } \frac{ 3\Im\varepsilon }{ |2\varepsilon+1|^2 } \;.
\end{aligned}
\end{equation}
We find that $S$ in Eq.~({\ref{def_s}}) is the quotient of the electric and chiral rates and amounts to:
\begin{equation}\label{sbulkabs1}
\begin{aligned}
S^{\text{lfc}} =  \frac{ 2 R_{IK} }{ 3 c |\bm d_{IK}|^2 } \left[ 2\Re\kappa - \frac{ \Im\kappa }{ \Im\varepsilon } \left( 2\Re\varepsilon +1\right)  \right]  \;.
\end{aligned}
\end{equation}

Furthermore, if we consider a bulk medium with $\kappa$ real-valued, we obtain 
\begin{equation}\label{sbulkabs2}
\begin{aligned}
S^{\text{lfc}} =  \frac{ 4 \kappa R_{IK} }{ 3 c |\bm d_{IK}|^2 } =  - \frac{2}{9} S^{(0)} \;,
\end{aligned}
\end{equation}
which is proportional to the uncorrected degree of discrimination found in Eq.~(\ref{squo}) and independent of the imaginary part of the permittivity. Lastly, we point out that Eqs.~(\ref{sbulkabs1}) and (\ref{sbulkabs2}) are again proportional to the ratio $R_{IK}/|\bm d_{IK}|^2$.

\subsection{Perfect chiral mirror}\label{secperfectchiralmirror}
\begin{figure}[htbp] 
\centering
\includegraphics[height=55mm,keepaspectratio]{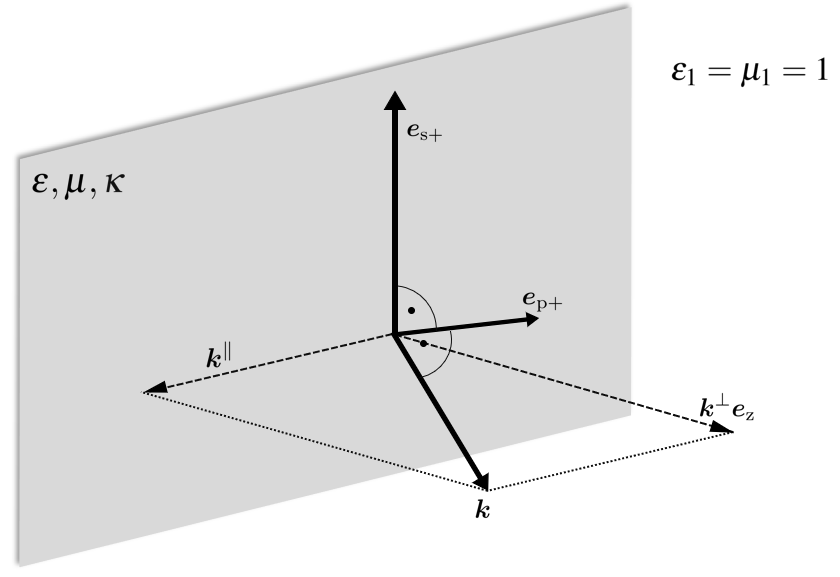}
\caption{Wave and polarization vectors of a plane wave, adapted from Ref.~\cite{SYB1}. The shaded area is to be understood as the boundary between the vacuum with permittivity $\varepsilon_1=1$ and permeability $\mu_1=1$, and the semi-infinite chiral medium with permeability $\varepsilon$, permeability $\mu$ and chiral cross-susceptibility $\kappa$. }
\label{hsgeo}
\end{figure}

Next, we examine a geometry of two semi-infinite half spaces separated by a planar interface. We take a material with chiral properties on one side, while the other half contains free space, see Fig.~\ref{hsgeo}. Therefore, the decay rate is the sum of the vacuum decay rate, as well as the electric and chiral contributions arising from the interaction of the molecule with this interface. We will now have a closer look at the chiral contribution. 

%
%
The substantial unit vectors for the reflected and transmitted waves are the following:

\begin{equation}
\begin{aligned}
\mathbf{e}_{s \pm}&=\mathbf{e}_{k \|} \times \mathbf{e}_{\it z} \;, \\
\mathbf{e}_{p \pm}&=\frac{1}{k}\left(k^{\|} \mathbf{e}_{\it z} \mp k^{\perp} \mathbf{e}_{k \|}\right),
\end{aligned}\label{evec}
\end{equation}
where $k^\parallel=\sqrt{k_x^2+k_y^2}$ and $k^\perp=\sqrt{k_0^2-k^{\parallel2}}$ such that  $\mathrm{Im}\,k^\perp>0$. As shown in Fig. \ref{hsgeo}, $k^\parallel$ and $k^\perp$ define the parallel and perpendicular components to the interface of the wave vector $\mathbf{k}$
associated to the reflected wave, whose magnitude is the wave number $k_0=\omega/c$. 

As a starting point, we take the general form for the scattering Green's tensor for a half space \cite{pchgreent}, given by:
\begin{equation}
\begin{aligned}
\mathbb{G}^{(1)}\left(\mathbf{r}, \mathbf{r}^{\prime}, \omega\right)=\frac{\mathrm{i}}{8 \pi^2} \int \frac{\mathrm{d}^2 k^{\|}}{k^{\perp}} \mathrm{e}^{\mathrm{i}\left[\mathbf{k}^{\|} \cdot\left(\mathbf{r}-\mathbf{r}^{\prime}\right)+k^{\perp}\left(z+z^{\prime}\right)\right]}&\\
\times\sum_{ \sigma= \mathrm{s,p} } \sum_{ \sigma' = \mathrm{s,p} } r_{ \sigma,\sigma' } \mathbf{e}_{\sigma +}\otimes\mathbf{e}_{\sigma' -} \;.
\end{aligned}\label{gchhs0}
\end{equation}

The reflection coefficients $r_{ \sigma,\sigma' }$ describe the proportion of reflected light with polarization $\sigma'$ for incident light with polarization $\sigma$. For chiral materials, these cross-terms do not vanish.

To gain insight, we start with an idealized medium that rotates the polarization of reflected light by $\pi/2$. This so-called perfect chiral mirror has the reflection coefficients $r_{sp},r_{ps}=\pm1$ and $r_{ss},r_{pp}=0$ \cite{pchgreent}. If the medium rotates the polarization of the incoming wave in clockwise direction, labelled \textquotedblleft right-handed\textquotedblright, the coefficients have the values $r_{sp}=1$ and $r_{ps}=-1$. In the contrary case of a \textquotedblleft left-handed\textquotedblright\, medium the polarization is rotated anti-clockwise and the coefficients have the values $r_{sp}=-1$ and $r_{ps}=1$. The scattering Green's tensor then takes the form
\begin{equation}
\begin{aligned}
\mathbb{G}^{(1)}\left(\mathbf{r}, \mathbf{r}^{\prime}, \omega\right)=\pm\frac{\mathrm{i}}{8 \pi^2} &\int \frac{\mathrm{d} ^2 k^{\|}}{k^{\perp}} \mathrm{e}^{\mathrm{i}\left[\mathbf{k}^{\|} \cdot\left(\mathbf{r}-\mathbf{r}^{\prime}\right)+k^{\perp}\left(z+z^{\prime}\right)\right]}&\\
&\times \left[\mathbf{e}_s\otimes\mathbf{e}_{p_-}-\mathbf{e}_{p_+}\otimes\mathbf{e}_s\right], \label{gchhs1}
\end{aligned}
\end{equation}
where $\pm$ denotes \textquotedblleft right-\textquotedblright\, and \textquotedblleft left-handed\textquotedblright\,  media, respectively. The result for the curl is
\begin{equation}
\begin{aligned}
\nabla\times\mathbb{G}^{(1)}=\mp\frac{k_0}{8 \pi^2}& \int\frac{\mathrm{d}^2 k^{\|}}{k^{\perp}} \mathrm{e}^{\mathrm{i}\left[\mathbf{k}^{\|} \cdot\left(\mathbf{r}-\mathbf{r}^{\prime}\right)+k^{\perp}\left(z+z^{\prime}\right)\right]}&\\
&\times \mathbf{e}_{k_+}\times\left[\mathbf{e}_s\otimes\mathbf{e}_{p_-}-\mathbf{e}_{p_+}\otimes\mathbf{e}_s\right],
\end{aligned}
\end{equation}
where $\mathbf{e}_{k}$ is the unit vector pointing in the direction of the reflected wave $\mathbf{k}$. 

%
%

Carrying out the cross product, taking the coincidence limit with $\bm r'\rightarrow \bm r$ and changing the integration variable, we find

\begin{equation}
\begin{aligned}
&\nabla\times\mathbb{G}^{(1)}\left(\mathbf{r}, \mathbf{r}, \omega\right)=\pm\frac{k_0}{8\pi}\int_0^{k_0} \mathrm{d} k^{\perp} \mathrm{e}^{2 \mathrm{i} k^{\perp} z} \\
& \times \left[\bigg(1-\frac{k^{\perp2}}{k_0^2}\bigg)(\mathbf{e}_{x}\otimes\mathbf{e}_{x}+\mathbf{e}_{y}\otimes\mathbf{e}_{y})+\frac{2k^{\|2}}{k_0^2}\mathbf{e}_{z}\otimes\mathbf{e}_{z}\right] \\
& \mp \frac{ \i k_0 }{ 8\pi }\int_0^{\infty} \mathrm{d} \kappa^{\perp} \mathrm{e}^{-2 \kappa^{\perp} z} \\
& \times \left[\bigg(1+\frac{\kappa^{\perp2}}{k_0^2}\bigg)(\mathbf{e}_{x}\otimes\mathbf{e}_{x}+\mathbf{e}_{y}\otimes\mathbf{e}_{y}) +\frac{2k^{\|2}}{k_0^2}\mathbf{e}_{z}\otimes\mathbf{e}_{z}\right] ,
\end{aligned}\label{ghsraw}
\end{equation}
where $\kappa^{\perp}=\sqrt{k_0^2+k^{\|2}}$. By splitting the integral in this way, we quickly identify the influence of travelling waves (the integral from 0 to $k_0$) and evanescent waves (the integral from 0 to $\infty$) and additionally, simplify the calculation of close or far limits.

\subsubsection{Retarded limit}
If the molecule is far from the interface, $z_{M}\gg c/\w$, an observer perceives an oscillating dipole that creates periodically changing fields. This is what we call the retarded limit. The rapidly oscillating integrand in Eq.~(\ref{ghsraw}) then leads to strong cancellations, leaving only a relevant contribution from the point with $k^\|=0$. We may therefore approximate
\begin{align}
k^\perp\simeq k_0\;. \label{ret approx mirror}
\end{align}

Solving the integrals by using these approximations, we discover the term of leading order 
\begin{align}
\nabla\times \mathbb{G}^{(1)}(\bm r, \bm r,\w)=\mp\frac{ \e^{2\i k_0 z} }{16\pi z^2}\myvec{1&0&0 \\ 0&1&0\\0&0&0}. \label{rpcm}
\end{align}

Substituting into Eq.~(\ref{chdr}), we discern the chiral contribution of a molecule at a distance of $z_M$ in front of a perfect chiral mirror in the retarded limit: 
\begin{equation}
\begin{aligned}
\Gamma_{ch}^{(1)}=\pm\frac{\mu_0\w_{IK}}{4\pi \hslash z_M^2} \sin\left(\frac{2\w_{IK} z_M}{c}\right) \text{Im}\left(\bm d_{IK}^{\|}\cdot \bm m_{IK}^{\|*}\right),
\end{aligned}\label{retmir0}
\end{equation}
where $\bm d_{IK}^\|, \bm m_{IK}^\|$ denote the components parallel to the interface for the respective dipoles. For isotropic molecules we may thus write 
\begin{equation}
\text{Im }\left(\bm d_{IK}^{\|}\cdot \bm m_{IK}^{\|*}\right)=\frac{2}{3}\text{Im }\left(\bm d_{IK}\cdot \bm m_{IK}^*\right)=\frac{2}{3}R_{IK},
\end{equation}
and thus 
\begin{equation}
\begin{aligned}
\Gamma_{ch}^{(1)}=&\pm\frac{\mu_0\w_{IK}R_{IK}}{6\pi \hslash z_M^2} \sin\left(\frac{2\w_{IK} z_M}{c}\right) \;.
\end{aligned}\label{retmir}
\end{equation} 
If both the mirror and the molecule are right-handed (positive sign and $R_{IK}>0$), we see a positive contribution to the decay rate (\ref{whole dr}). If the mirror and the molecule are left-handed (negative sign and $R_{IK}<0$), it is also positive. Only if the handedness is dissimilar, there is negative contribution. This behavior is the same as that found for the chiral bulk medium in Eq.~(\ref{chdrbulk}). For comparison, the analogous electric contribution in front of an idealized dielectric mirror with $r_{ss},r_{pp}=1$ and $r_{sp},r_{ps}=0$ reads directly from Eq.~(\ref{ddr}) as follows:
\begin{equation}
\begin{aligned}
\Gamma_{el}^{(1)}=\frac{\mu_0  \big| \bm d_{IK}^{\|}\big|^{2} }{8\pi\hslash z_M } \sin\left(\frac{2\w_{IK} z_M}{c}\right) \;.\label{drreths}
\end{aligned}
\end{equation}

For the chiral contribution (\ref{retmir}), the distance is squared. Therefore, its influence to the total decay rate (\ref{whole dr}) is smaller than the electric counterpart, even though the transition frequency affects it directly. The oscillatory structure remains the same in both cases. In general, the chiral interaction is smaller due to lower magnitudes of the magnetic dipole. Although we appreciate the same vectorial structure encoded in the optical rotatory strength as in the chiral bulk medium (\ref{chdrbulk}), in the current case only the components of the electric and magnetic dipoles that are parallel to the interface contribute. This means that the chiral effect is greatest when the parallel electric and magnetic dipoles are colinear too.  

The degree of discrimination does not lead to the quotient of the specific electric and chiral contributions of the geometry this time, as the non-discriminating rate $\Gamma_{nd}$ consists of the vacuum rate and the scattering part in Eq.~(\ref{drreths}). Using the definition in Eq.~(\ref{def_s}), we now obtain:
\begin{equation} \label{schmir}
\begin{aligned}
S^{(1)} &= \frac{\Gamma_{ch}^{(1)}}{\Gamma^{(0)} } \simeq \pm \frac{ c R_{IK} }{2 \w_{IK}^2 z_M^2 |\bm d_{IK}|^2} \sin\left(\frac{2 \w_{IK} z_M}{c}\right). 
\end{aligned}
\end{equation}

This result shows that at distances far away from the interface the degree of discrimination rapidly decreases to zero. There is however a clear distinction between molecules with same and opposite handedness, expressed in the sign. While the decay rates in this limit depend on the parts of the dipole that are parallel to the surface, Eq.~(\ref{schmir}) is determined by the whole dipole, the same way the other degrees of discrimination were in Sec. \ref{secchiralbulk}. A recurring factor in all of them is the quotient $R_{IK}/|\bm d_{IK}|^2$.

\subsubsection{Nonretarded limit}
The nonretarded limit is the opposite case where the molecule is very close to the interface $z_M \ll c/ \omega$. The dipole can now be regarded as quasi-static, since any fields reflected by the mirror arrive at the atom instantly. The integrals (\ref{ghsraw}) have their main contribution where $k^\|$ is large. Letting $k^\|\rightarrow\infty$, we may approximate
\begin{align}
k^\perp\simeq \i k^\|\;. \label{nret approx mirror}
\end{align}
The curl of the scattering Green's tensor then reads
\begin{equation}
\nabla\times\mathbb{G}^{(1)}(\boldsymbol{r}, \boldsymbol{r}, \omega)=\mp\frac{\i }{32\pi k_0 z^3}\myvec{1&0&0 \\ 0&1&0\\0&0&2 }. \label{nrpcm}
\end{equation}
The chiral contribution of a molecule in a nonretarded distance to a perfect chiral mirror follows straightforwardly:
\begin{equation}
\begin{aligned}
\Gamma_{ch}^{(1)}=\pm\frac{\mu_0c}{8\hslash\pi z_M^3}\text{Im}\left(\bm d_{IK}\cdot \bm m_{IK}^*+d_{z}m_{z}^*\right).
\end{aligned}\label{nretmir0}
\end{equation}
If we once again assume that the dipoles are isotropic, we may write
\begin{equation}
\text{Im }\left(\bm d_{IK}\cdot \bm m_{IK}^*+d_{z}m_{z}^*\right)=\frac{4}{3}\text{Im }\left(\bm d_{IK}\cdot \bm m_{IK}^*\right)=\frac{4}{3}R_{IK}.
\end{equation}
The chiral contribution for isotropic molecules reads
\begin{equation}
\begin{aligned}
\Gamma_{ch}^{(1)}=\pm\frac{\mu_0cR_{IK}}{8\hslash\pi z_M^3}\;.
\end{aligned}\label{nretmir}
\end{equation}

Correspondingly, we discover again that matching handedness enhances and opposite handedness diminishes the total decay rate (\ref{whole dr}). Between the retarded or nonretarded limits, the latter has a far greater magnitude. This is owed to the cubed inverse distance dependence in a low distance limit. It is thus more apt for detection and/or for enantiomer discrimination. By contrast, the electric contribution for an atom in front of a perfect dielectric mirror, whose reflection coefficients are $r_{ss}, r_{pp}=1$, is vanishing in the nonretarded limit \cite{SYB1,SYB2}. What makes this result even more remarkable is the scattering contribution being purely chiral. The degree of discrimination as introduced in the previous section therefore results in a quotient with only the vacuum decay rate, as in Eq.~(\ref{schmir}):
\begin{equation}\label{schmirnret}
\begin{aligned}
S^{(1)} = \frac{ \Gamma_{ch}^{(1)} }{ \Gamma^{(0)} } = \pm \frac{ 3c^2 R_{IK} }{  \w_{IK}^3 z_M^3 |\bm d_{IK}|^2 } \;.
\end{aligned}
\end{equation}
In this limit, Eq.~(\ref{schmirnret}) shows an increase the closer the molecule is to the mirror, dictating a noteworthy discrimination between the enantiomers through the opposite signs they have. The magnitude is modulated by the inverse frequency and the factor $R_{IK}/|\bm d_{IK}|^2$ for each specific molecule.


\subsection{Half space}\label{Sec HS}
Having discussed the idealised chiral mirror, we now generalize to a more realistic medium with diagonal reflection coefficients $r_{ss},r_{pp}\neq0$. The crossed reflection coefficients obey the relation $r_{sp}=-r_{ps}$ due to reciprocity \cite{marachevsky}. For conciseness, we split the scattering Green's tensor in Eq.~(\ref{gchhs0}) into two parts, perform the curl and find
\begin{equation}
\begin{aligned}
&\nabla\times\mathbb G^{(1)}\left(\mathbf{r}, \mathbf{r}, \omega\right) = \frac{k_0}{8 \pi^2} \int\frac{\mathrm{d}^2 k^{\|}}{k^{\perp}} \mathrm{e}^{2\mathrm{i}k^{\perp}z} \left[ \mathbf{e}_{s}\otimes\mathbf{e}_{p_-} r_{pp} \right.\\
&\left. -\mathbf{e}_{p_+}\otimes\mathbf{e}_{s}r_{ss} + \left(\mathbf{e}_{p_+}\otimes\mathbf{e}_{p_-}+\mathbf{e}_s\otimes\mathbf{e}_s\right) r_{sp} \right].
\end{aligned}\label{gichhs}
\end{equation}

To make an assessment of the chiral contribution we must find the imaginary part of the scattering Green's tensor, which is dependent on the reflection coefficients, which in general have real and imaginary parts. To complete the analysis, we need to calculate the reflection coefficients in the limits and substitute the imaginary part into the decay rate.

\subsubsection{Retarded limit}


The scattering Green's tensor has no analytical solution without further approximations due to the reflection Fresnel coefficients dependency on $k^\parallel$. In the retarded limit ($\omega z_M/c\gg1$), the main contribution comes from the region with small $k_\parallel$. We find Fresnel coefficients for an interface constituted by an isotropic chiral medium and vacuum: 
\begin{equation}
\begin{aligned}
r_{ sp } &= \frac{ 2 \mathrm{i}\, ( k_{+}|k_{-}| - k_{-}|k_{+}| ) }{ 2\left(\frac{ \varepsilon + \mu }{n_r}\right)(k_{+}k_{-} + |k_{-}k_{+}|) + ( k_{+}|k_{-}| + k_{-}|k_{+}| ) } ,\\\\\
r_{ ps } &= - r_{sp} \,, \\
r_{ ss } &= \frac{ 2( k_{+}k_{-} - |k_{-}k_{+}| ) - \left(\frac{ \varepsilon - \mu }{n_r}\right)(k_{+}|k_{-}| + k_{-}|k_{+}|) }{ 2(k_{+}k_{-} + |k_{-}k_{+}|) + \left(\frac{ \varepsilon + \mu }{n_r}\right)( k_{+}|k_{-}| + k_{-}|k_{+}| ) } , \\\\\
r_{ \mathrm{pp} } &= \frac{ 2( k_{+}k_{-} - |k_{-}k_{+}| ) + \left(\frac{ \varepsilon - \mu }{n_r}\right)(k_{+}|k_{-}| + k_{-}|k_{+}|) }{ 2(k_{+}k_{-} + |k_{-}k_{+}|) + \left(\frac{ \varepsilon + \mu }{n_r}\right)( k_{+}|k_{-}| + k_{-}|k_{+}| ) } ,\\\
\end{aligned}
\end{equation}
where we recall the definitions for $k_-$ and $k_+$ given in Eqs.~(\ref{cwv}). To obtain these coefficients, the approximations $k^{\|}=0$ and 
\begin{equation}\label{ret approx hs}
k^\perp\simeq n_r k_0 =\sqrt{\varepsilon(\w)\mu(\w)} k_0
\end{equation}
were applied to their general expressions given in Ref.~\cite{refcof}, and can be found in Appendix \ref{C}. In this limit, we found that for $\kappa\in\mathbb{C}$, $r_{ sp }$ will not vanish. For this reason, we consider the more general case of $\varepsilon,\kappa\in\mathbb{C}$ to analyze the chiral effects  in this subsection.

The curl of the scattering Green's tensor in the retarded limit can be calculated with the approximations (\ref{ret approx mirror}) and (\ref{ret approx hs}), when calculating the integrals in Eq.~(\ref{gichhs}) and yields: 
\begin{equation}
\begin{aligned}
&\nabla\times\text{Im }\mathbb{G}^{(1)}(\boldsymbol{r}, \boldsymbol{r}, \omega)=\frac{k_0}{8\pi z} \left\{ \left( \mathbf{e}_{x}\otimes\mathbf{e}_{x}+\mathbf{e}_{y}\otimes\mathbf{e}_{y} \right) \right. \\
& \times \left[ \left(1-\cos(k_0 z_M)\right) \Re(r_{sp}) -\sin(k_0 z_M)\Im(r_{sp}) \right] \\
& +\frac12 \left( \mathbf{e}_{x}\otimes\mathbf{e}_{y}-\mathbf{e}_{y}\otimes\mathbf{e}_{x} \right)\left[ \sin(k_0 z_M)\Im(r_{ss} + r_{pp}) \right. \\
& \left. \left. -\cos(k_0 z_M)\Re(r_{ss}+r_{pp}) \right] \right\} .
\end{aligned}\label{cghsret}
\end{equation}

The chiral contribution of a molecule in front of a chiral half space follows by substituting Eq.~(\ref{cghsret}) into Eq.~(\ref{chdr}):
\begin{equation}
\begin{aligned}
&\Gamma_{ch}^{(1)}=\frac{\mu_0 \w_{IK}^2}{2\pi \hslash c z_M} \left\{ \Im \big(\bm d_{IK}^\| \cdot \bm m_{IK}^{*\|}\big) \right. \\
&\times\left[(\cos(\tfrac{2\w_{IK} z_M}{c})-1)\Re(r_{sp})+\sin(\tfrac{2\w_{IK} z_M}{c})\Im(r_{sp}) \right] \\
& +\frac12 \,\Im\left[\left(\bm d_{IK}\times\bm m_{IK}^*\right)\cdot\mathbf{e}_{z}\right] \left[ \cos(\tfrac{2\w_{IK} z_M}{c})\Re(r_{ss}+r_{pp}) \right. \\
& \left. \left. -\sin(\tfrac{2\w_{IK} z_M}{c})\Im(r_{ss}+r_{pp})\right] \right\}, 
\end{aligned}\label{chhsret}
\end{equation}
where the first term is a discriminating chiral contribution, while the second term, albeit dependent of $\kappa$, does not distinguish between handedness.
As expected, the vectorial dependency is the same as in the perfect chiral mirror (\ref{retmir0}), but is now modulated by the Fresnel coefficient $r_{sp}$. On the other hand, the terms with diagonal reflection were absent in the case of a perfect chiral mirror. Their vectorial structure will contribute maximally when the unit vector $\mathbf{e}_{\it z}$, the electric and magnetic dipoles form an orthogonal triad. Interestingly, they vanish if the molecule is isotropic, in which case Eq.~(\ref{chhsret}) reduces to the purely chiral contribution:
\begin{equation}
\begin{aligned}
&\Gamma_{ch}^{(1)}=\frac{\mu_0\w_{IK}^2}{3\pi c \hslash z_M}R_{IK} \\
&\times \left[(\cos(\tfrac{2\w_{IK} z_M}{c})-1)\Re(r_{sp})+\sin(\tfrac{2\w_{IK} z_M}{c})\Im(r_{sp}) \right] .
\end{aligned}
\end{equation}
As seen with the idealised mirror, there is an oscillating behavior of the chiral contribution in the retarded limit. Astoundingly, the distance dependence of $1/z_M$ is more pronounced than for the mirror (because in the retarded limit the distances are large, resulting in $1/z_M>1/z^2_M$) making the chiral effect larger in magnitude for a more realistic material. Ref.~\cite{Guzatov1} presents numerical results to study a wider range of distances by solving the complicated integrals over $k^\perp$ or $\kappa^\perp$ in Eq.~(\ref{ghsraw}) or $h$ or $p$ in the notation of that article. A detailed discussion exhibiting the differences between the results of Ref.~\cite{Guzatov1} and ours will be provided in the next section when the nonretarded limit is investigated. 

The scattering electric contribution of an atom in a great distance to a dielectric half space is given by \cite{SYB2} 
\begin{eqnarray}
\Gamma_{el}^{(1)} &=& \frac{\mu_0\w_{IK}^2}{4\hslash\pi z_M} \big| \bm d_{IK}^{\|}\big|^{2}  \left[\cos(\tfrac{2\w_{IK} z_M}{c})\Im(r_{s}) \right. \nonumber\\
&& \left. + \sin(\tfrac{2\w_{IK}z_M}{c}) \Re(r_{s}) \right] \;,
\end{eqnarray}
with 
\begin{equation}
r_{s} = \frac{\sqrt{\mu}-\sqrt{\varepsilon}}{\sqrt\mu+\sqrt\varepsilon} \;.
\end{equation}

As seen before with the bulk medium and perfect mirror, the chiral contribution for left- and right-handed media differs only in the sign of the expression, rendering the degree of discrimination to be essentially the quotient of chiral and electric contributions as seen in Eq.~(\ref{squo}). It has the value
\begin{equation}\label{schhsret}
\begin{aligned}
& S^{(1)} =  \frac{ \Gamma_{ch}^{(1)} }{\Gamma^{(0)}+\Gamma_{el}^{(1)} } \;, \\
&=  \frac{ R_{IK}  }{ \w_{IK} z_M |\bm d_{IK}|^2 } \bigg[(\cos(\tfrac{2\w_{IK} z_M}{c})-1)\Re(r_{sp}) \\
& \quad +\sin(\tfrac{2\w_{IK} z_M}{c})\Im(r_{sp}) \bigg] .
\end{aligned}
\end{equation}
This degree of discrimination is surprisingly only modulated by the crossed Fresnel coefficient $r_{sp}$ and shares the quotient $R_{IK}/|\bm d_{IK}|^2$ as global factor with the previous degrees of discrimination. Except for the clear difference of complex material properties $\varepsilon,\kappa\in\mathbb{C}$, we furthermore observe that this result is an oscillating function that decreases with the molecule's position $z_M$ but slower in comparison with Eq.~(\ref{schmir}) for the chiral mirror. 

Note that our plate-assisted Purcell rates exhibit spatial oscillations as a result of interference between photons emited towards and then reflected from the plate. In contrast, chiral discriminatory energy transfer rates \cite{JanineNJP,Craig-Thirunamachandran} between an excited donor molecule and an absorbing acceptor molecule are purely monotonous.

\subsubsection{Nonretarded limit}


Lastly, we consider the nonretarded limit $z_M\ll c/\w$. In this regime we can apply the approximation (\ref{nret approx mirror}) together with 
\begin{align}
k^\perp\simeq \i k^\|. \label{nret approx hs}
\end{align}

The reflection coefficients of an isotropic chiral medium that fills the lower half space are deduced in Ref.~\cite{pchgreent}, and given here by Eqs.~(\ref{rsp})-(\ref{r_pp}). As a result we thus obtain
\begin{equation}
\begin{aligned}
r_{sp} &=-r_{ps}=\frac{2 \mathrm{i} \kappa}{\varepsilon \mu-\kappa^2+\varepsilon+\mu+1}, \\
r_{ss}&=\frac{\varepsilon \mu-\kappa^2-\varepsilon+\mu-1}{\varepsilon \mu-\kappa^2+\varepsilon+\mu+1}, \\
r_{pp}&=\frac{\varepsilon \mu-\kappa^2+\varepsilon-\mu-1}{\varepsilon \mu-\kappa^2+\varepsilon+\mu+1} .
\end{aligned}
\end{equation}
For simplicity, we choose lossless media in this case, corresponding to real $\varepsilon$, $\mu$ and $\kappa$. Here, we may verify that for this approximation the reflection coefficients $r_{ss}, r_{pp}$ are indeed real and $r_{sp}$ is purely imaginary. Using these reflection coefficients and calculating the integrals with the approximations (\ref{nret approx mirror}) and (\ref{nret approx hs}) in the same way as for the idealized mirror, we obtain the curl of the imaginary scattering Green's tensor anew: 

\begin{multline}
\nabla\times\text{Im }\mathbb{G}^{(1)}(\boldsymbol{r}, \boldsymbol{r}, \omega)=\\
\frac{1}{16\pi z^2}\myvec{\i r_{sp}-(r_{pp}+r_{ss})zk_0&0 \\ (r_{pp}+r_{ss})zk_0&\i r_{sp}&0\\0&0& 2\i r_{sp}}.
\end{multline}

Substituting this into (\ref{chdr}), we discover the chiral contribution of a particle in a nonretarded distance to an isotropic chiral interface:
\begin{equation}
\begin{aligned}
\Gamma_{ch}^{(1)}&=-\frac{\i\mu_0\w_{IK}}{4\pi\hslash z^2}r_{sp}\text{Im }(\bm d_{IK}\cdot\bm m_{IK}^*+d_{IK}^{z}m_{IK}^{z*})\\
-&\frac{\mu_0\w_{IK}^2}{4\pi c\hslash z}(r_{pp}+r_{ss})\Im \left[\left(\bm d_{IK}\times \bm m_{IK}^*\right)\cdot \mathbf{e}_{\it z}\right] .
\end{aligned}\label{chhsnret}
\end{equation}

In comparison with the perfect mirror (\ref{nretmir}), we find strong resemblance for the terms with cross-reflection. Nevertheless, the distance dependence is now weakened with only $1/z_M^2$ instead of $1/z_M^3$, resulting in smaller decay rates. Additionally, the reflection-coefficient scaling of this contribution is now dependent of material properties. The terms with diagonal reflection were absent in the case of a perfect chiral mirror. They feature an even weaker distance dependence of $1/z_M$ and share the same vectorial structure as in the retarded limit. Therefore, the electric and magnetic dipoles will contribute maximally again when define an orthogonal 
triad with the unit vector $\mathbf{e}_{\it z}$. Using the optical rotatory strength and assuming the molecule to be isotropic, the chiral contribution has the compact form
\begin{equation}
\begin{aligned}
\Gamma_{ch}^{(1)} =-\frac{\i \w_{IK}R_{IK}}{3\pi\hslash \varepsilon_0 c^2 z_M^2}r_{sp}.
\end{aligned}\label{drchhsiso}
\end{equation}

Recall that the total decay rate is the sum of the electric and chiral contributions. As a consistency check, if we set any of the chiral properties, namely the chiral cross-susceptibility $\kappa$, the crossed reflection coefficients $r_{sp},r_{ps}$ or the optical rotatory strength $R_{IK}$ to 0, we find vanishing results for all the chiral decay rates calculated in this section recovering the well-known results for the purely electric contributions. In agreement with Curie's principle, both interacting parties need to have chiral properties to create chirality-induced effects. At this point, it is important to compare our results with the semiclassical approach of Ref.~\cite{Guzatov1}. First, a notable difference between Eq.~(20) of Ref.~\cite{Guzatov1} and our Eqs.~(\ref{chhsnret}) and (\ref{drchhsiso}) is that ours have a $1/z_{\mathrm{M}}^2$  distance dependence in the chiral sensitive part while the former $1/z_{\mathrm{M}}^3$. Second, another big difference is that the chiral response of the molecule in Eq.~(20) of Ref.~\cite{Guzatov1} is dictated by $\mathrm{Re}(\mathbf{d}\cdot\mathbf{m}^*+d_z m^*_z)$ instead of the optical rotatory strength $R_{\mathrm{IK}}$ defined by the imaginary part of the same quantities, as our Eqs.~(\ref{chhsnret}) and (\ref{drchhsiso}) show. Third, Eq.~(20) of Ref.~\cite{Guzatov1} vanishes for lossless chiral media with $\mathrm{Im}(\varepsilon)=\mathrm{Im}(\mu)=0$, which differs from our results of Secs.~\ref{secperfectchiralmirror} and this one. Then, the response interpretation can be straightforwardly applied to our Eqs.~(\ref{chhsnret}) and (\ref{drchhsiso}) where one can appreciate how only the mixed polarisations $\mathrm{s}$ and $\mathrm{p}$ via the Fresnel coefficient $r_{\mathrm{sp}}$ couple to the optical rotatory strength $R_{\mathrm{IK}}$ of the molecule agreeing with Curie's principle. Nevertheless, this interpretation might be difficult to read from Eq.~(20) of Ref.~\cite{Guzatov1} because the electromagnetic field is determined by matching the boundary conditions at the interface as seen in its Sec.~2, so the Fresnel coefficients displayed in Appendix \ref{C} associated to the material response could not be easily extracted. These discrepancies could originate from our use of the Boys-Post constitutive relationships (\ref{constrel}) instead of the Drude-Born-Fedorov ones employed in Ref.~\cite{Guzatov1}. In fact, once a set of constitutive relationships is chosen, the differences are immediate. This can be directly appreciated in the wavenumbers $k_{\mathrm{L}}$ and $k_{\mathrm{R}}$ obtained after performing Bohren's decomposition \cite{Bohren1,Bohren2} corresponding to the Drude--Born--Fedorov set given by Eq.~(4) of Ref.~\cite{Guzatov1} or Eqs.~(57) of Ref.~\cite{Klimov2} for bi-isotropic chiral materials and Eq.~(2) of Refs.~\cite{Bohren1,Bohren2} for an isotropic chiral medium. In our case, the wave numbers are given by Eqs.~(\ref{cwv}), which were taken from Eqs.~(27) of Ref.~\cite{refcof} and Eqs.~(16) of Ref.~\cite{chbulk}. In this way, the two choices lead to very distinct results that are difficult to compare.

Let us now compare the magnitude of the chiral contribution to the electric one \cite{SYB2} 
\begin{equation}
\begin{aligned}
\Gamma_{el}^{(1)} & = \frac{ |\bm d_{IK}|^2+(\bm d_{IK}\cdot\mathbf e_z)^2}{8\hslash\pi\varepsilon_0\w_{IK}^2z_M^3}\Im \left( r_p \right)\, , \\
&= \frac{|\bm d_{IK}|^2+(\bm d_{IK}\cdot\mathbf e_z)^2}{8\hslash\pi\varepsilon_0\w_{IK}^2z_M^3}\frac{\Im\varepsilon}{|\varepsilon+1|} \;,
\end{aligned}
\end{equation}
which notably requires the permittivity $\varepsilon$ to have a non-zero imaginary part so as not to vanish, i.e., the medium must be lossy. In contrast to the chiral contribution (\ref{drchhsiso}), however, this requirement is not necessary because we derived it by assuming only real $\varepsilon$, $\mu$ and $\kappa$ as mentioned before in this section. This is a significant result because, as happened with the perfect chiral mirror, it means that the whole scattering contributions to the decay rate (\ref{whole dr}) in this limit are purely chiral too. 

We close this section by presenting the corresponding degree of discrimination for this configuration, which is
\begin{equation}\label{schhsnret}
\begin{aligned}
S^{(1)} = \frac{ \Gamma_{ch}^{(1)} }{ \Gamma^{(0)} } 
= \frac{ \i c R_{IK}\, r_{sp} }{ \w_{IK}^2 z_M^2 |\bm d_{IK}|^2 } \;.
\end{aligned}
\end{equation}
Apart from the usual quotient $R_{IK}/|\bm d_{IK}|^2$ that regulates the magnitude of this quantity, it is interesting to see that only the crossed Fresnel coefficient $r_{sp}$ governs this quantity in a similar way as found in the retarded degree of discrimination (\ref{schhsret}). In comparison with Eq.~(\ref{schmirnret}), we appreciate that $S^{(1)}$ given by Eq.~(\ref{schhsnret}) increases slower than the former when the molecule's position $z_M$ approaches the interface.

\section{Conclusions} \label{secsum}

The goal of this work was to extend the theory of macroscopic QED to evaluate the chiral Purcell effect. From Fermi's Golden Rule, we derived an expression for the chiral decay rate of molecules which is dependent of the Green's tensor. Taking a set of constitutive relations that account for the chirality of the environment, we were able to calculate the chiral contribution for different geometries. To amount for the chirality of a molecule, we included the magnetic-dipole transition matrix in the derivation of a general expression for the chiral contribution. The total decay rate is the sum of the electric and chiral contributions.

The chiral contribution in a bulk medium is very similar in appearance to the electric contribution, only having the optical rotatory strength of the molecule and the chiral cross-susceptibility of the medium as factors instead of the electric dipole transition matrix elements. The calculated local-field correction factor also resembles the electric version, but with an additional function of only the permittivity and the permeability. The chiral contribution of an absorbing medium with local-field corrections was also analyzed, uncovering an intricate function of the real and imaginary parts of the permittivity, permeability and the chiral cross-susceptibility of the medium. This quantity shares the dependence on the inverse of the cube radius of the Onsager cavity with its electric analog. An idealised chiral mirror was considered to gain insight on the nature of chiral effects above a chiral half space. In the retarded limit, the chiral contribution exhibits an oscillating structure similar to the electric case, but the effect is weaker due to a greater inverse distance dependence. Moreover, the chiral contribution in the nonretarded limit is purely non-electric because the electric contribution vanishes.
It is also greater in magnitude than in the retarded limit.

When considering a chiral medium with conceivable properties in the retarded limit, we found that $\kappa$ is required to be complex to yield nonzero results for the reflection coefficients. The corresponding chiral contribution has an additional purely anisotropic term that is dependent on the diagonal Fresnel coefficients, but otherwise keeps the structure found for the mirror with a greater inverse distance dependence. For the half space in the nonretarded limit, it was possible to derive results using real $\varepsilon, \mu, \kappa$. The magnitude of the corresponding chiral contribution dropped in contrast with the ideal mirror and a weak anisotropic term as seen in the retarded limit was found. In both limits the decay rates are dependent on the optical rotatory strength $R_{IK}$ of the molecule and the chiral cross-susceptibility $\kappa$ of the medium.

\begin{figure}[htbp] 
\centering
\begin{tikzpicture}
\node at (0, 3) [minimum width=3cm, minimum height=3cm, align=center] {General};
\node at (3, 3) [minimum width=3cm, minimum height=3cm, align=center] {\includegraphics[width=2.4cm, keepaspectratio]{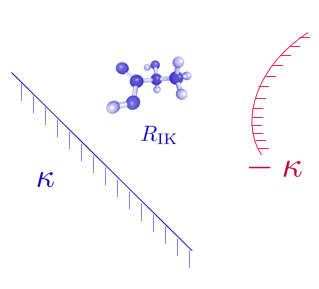}};
\node at (6, 3) [minimum width=3cm, minimum height=3cm, align=center] {Decay depends\\ on relative \\ handedness \\ \\ $\Gamma_{ch}\propto R_{IK}/|\bm d_{0K}|^2$ };

\node at (0, 1) [minimum width=3cm, minimum height=3cm, align=center] {Bulk};
\node at (3, 1) [minimum width=3cm, minimum height=3cm, align=center] {\includegraphics[width=2.35cm, keepaspectratio]{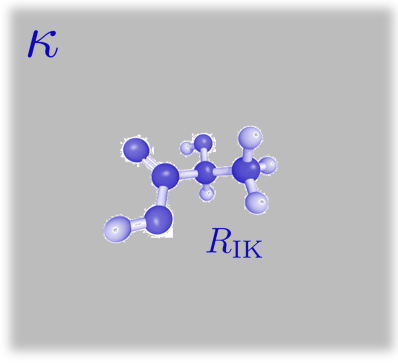}};
\node at (6, 1) [minimum width=3cm, minimum height=3cm, align=center] {$\Gamma_{ch}\propto \kappa R_{IK}$};

\node at (0, -1) [minimum width=3cm, minimum height=3cm, align=center] {Local-field,\\ absorbing media};
\node at (3, -1) [minimum width=3cm, minimum height=3cm, align=center] {\includegraphics[width=2.35cm, keepaspectratio]{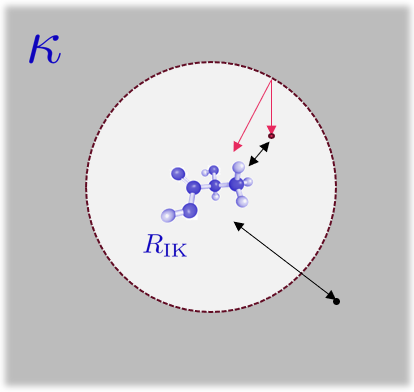}};
\node at (6, -1) [minimum width=3cm, minimum height=3cm, align=center] {$\Gamma_{ch}\propto R_{IK}$ \\ $\times\big[ 2\Re\kappa \Im \varepsilon $\\ $-\Im\kappa\big( 2\Re\varepsilon +1\big) \big]$};
\node at (0, -3) [minimum width=3cm, minimum height=3cm, align=center] {Half space};
\node at (3, -3) [minimum width=3cm, minimum height=3cm, align=center] {\includegraphics[height=2.4cm, keepaspectratio]{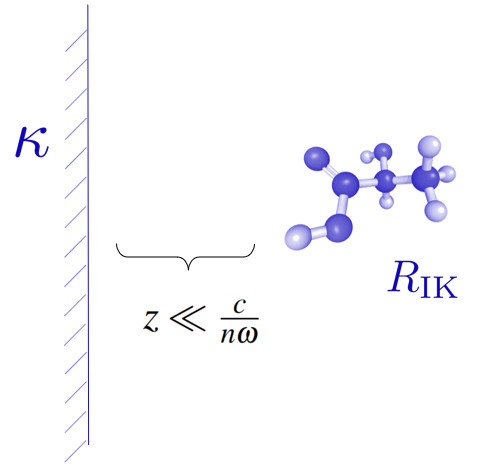}};
\node at (6, -3) [minimum width=3cm, minimum height=3cm, align=center] {$|\Gamma_{ch}^{\text{ret}}|<|\Gamma_{ch}^{\text{nret}}|$ \\  \\ anisotropic \\ effects};

\end{tikzpicture}
\caption{Most important results, visualized with propylene oxide as chiral molecule.}
\label{concfig}
\end{figure}

A few general results are the following: in all considered cases, the chiral bulk, the mirror and half space the sign of the chiral rate depends on the relative sign of the optical rotation of the medium and molecule (for an overview, see Fig.~\ref{concfig}). For the bulk medium, the rate is enhanced for coinciding handedness of molecule and medium. In other cases, enhancement or reduction depends on the relative phases of the medium response function and, for the half space, also on the distance. This is reflected in the equations with the signs of $R_{IK}$ and $\kappa$, and it is a result that we expected due to Curie's principle. In a bulk medium, the chiral effect is about 6 hyperfine constants smaller in magnitude than the electric effect. For the half space, it is apparent that the chiral interaction is most effective at very small distances and for isotropic molecules. Some smaller 
anisotropic effects that rely on diagonal reflection can be found. 
All degrees of discrimination considered in this work depend on the ratio $R_{IK}/|\bm d_{IK}|^2$.

Founded on this work, a logical next step would be to analyze further geometries, such as the vicinity of a spherical surface or inside a spherical cavity of finite size. This work can also be extended to a more general class of materials, it would be interesting to calculate the decay rates for nonlocal and nonreciprocal media. This and further theoretical work can lay the foundation for future experimental work towards observing enentiosensitive rates. One such scenario could be traversing of a tightly focused beam of excited chiral molecules \cite{Stickler} through a chiral cavity \cite{Voronin}, followed by state detection.

\newpage
\section{Acknowledgments}
This work has been supported by the German Research Foundation (DFG), Project No. 328961117 -- SFB 1319 ELCH. O. J. F. has been supported by the postdoctoral fellowship CONACYT-800966.

\appendix
\section{Vector wave functions $\bm{V}$ and $\bm{W}$} \label{A}

\renewcommand{\theequation}{A.\arabic{equation}}
\setcounter{equation}{0}
The vector wave functions $\bm V$ and $\bm W$ are defined as in article \cite{chbulk}:
\begin{equation}
\begin{aligned}
&\bm V_{\eo  \mathit{mn} }(\bm r ,k_0 )
 =  \frac{1}{\sqrt{2}}\left[ \bm M_{\eo \mathit{mn} }(\bm r  )+\bm N_{\eo \mathit{mn} }(\bm r  ) \right] \,, \\
& = 
\frac{1}{\sqrt{2}}
\left[
\mp \frac{m}{\sin \theta } P_{\mathit{mn}}(\cos\theta) 
\begin{array}{l}
\sin  \\ \cos 
\end{array}\!\! (m \phi ) \right.\\
& \times \left( 
j_{\mathit{n}} (  k_0  r ) \mathbf e_{\theta} 
+ 
\frac{1}{ k_0  r} \frac{\partial}{\partial  r } [ r j_{\mathit{n}} ( k_0r)]\mathbf e _\phi 
\right)
\\
&+ \frac{\partial P_{\mathit{nm}} ( \cos \theta  )}{\partial \theta}
\begin{array}{l}
\cos  \\ \sin 
\end{array} \!\! (m \phi ) \\ 
&
\times \bigg( 
\frac{1}{ k_0  r} \frac{\partial}{\partial  r } [ r j_{\mathit{n}} ( k_0r)]\mathbf e _\theta
- 
j_{\mathit{n}} (  k_0  r ) \mathbf e_{\phi}
\bigg) 
\\ 
& \left.
+ 
n ( n+1) P_{\mathit{nm}} ( \cos \theta  )
\begin{array}{l}
\cos  \\ \sin 
\end{array}\!\! (m \phi ) 
\frac{j_{\mathit{n}} ( k_0 r )}{k r} \mathbf e_r 
\right] . 
\end{aligned}
\end{equation}


\begin{equation}
\begin{aligned}
& \bm W_{\eo  \mathit{mn}}(\bm r, k_0  )
 =  \frac{ 1 }{\sqrt{2}} \left[ \bm M_{\eo \mathit{mn}}(\bm r  )-\bm N_{\eo \mathit{mn}}(\bm r  ) \right] \,, \\
& = 
\frac{1}{\sqrt{2}}
\Bigg[
\mp \frac{m}{\sin  \theta } P_{\mathit{nm}}(\cos\theta) 
\begin{array}{l}
\sin  \\ \cos 
\end{array}\!\! (m \phi ) \\
&
\times \bigg( 
j_{\mathit{n}} (  k_0  r ) \mathbf e_{\theta} 
-
\frac{1}{ k_0  r} \frac{\partial}{\partial  r } [ r j_{\mathit{n}} ( k_0 r)]\mathbf e _\phi 
\bigg)
\\
& 
- 
\frac{\partial P_{\mathit{nm}} ( \cos \theta )}{\partial \theta}
\begin{array}{l}
\cos  \\ \sin 
\end{array} \!\! (m \phi ) \\
&
\times 
\bigg( 
\frac{1}{ k_0  r} \frac{\partial}{\partial  r } [ r j_{\mathit{n}} ( k_0r)]\mathbf e _\theta
+ 
j_{\mathit{n}} (  k_0  r ) \mathbf e_{\phi}
\bigg) 
\\
& 
\left. - 
n ( n+1) P_{\mathit{nm}} ( \cos\theta  )
\begin{array}{l}
\cos  \\ \sin 
\end{array}\!\! (m \phi ) 
\frac{j_{\mathit{n}} ( k_0 r )}{k_0 r} \mathbf e_r 
\right].
\end{aligned} 
\end{equation}

The superscript \textquotedblleft(1)" in Eq.~(\ref{chbulk1}) indicates that the spherical Bessel function $j_{\mathit{n}}$ should be replaced by the spherical Hankel function $h_{\mathit{n}}^{(1)}$, in congruence with the cited article.



\section{Local-field correction for decay}\label{B}
\renewcommand{\theequation}{B.\arabic{equation}}
\setcounter{equation}{0}

\begin{figure}
\centering
\includegraphics[width=0.45\textwidth]{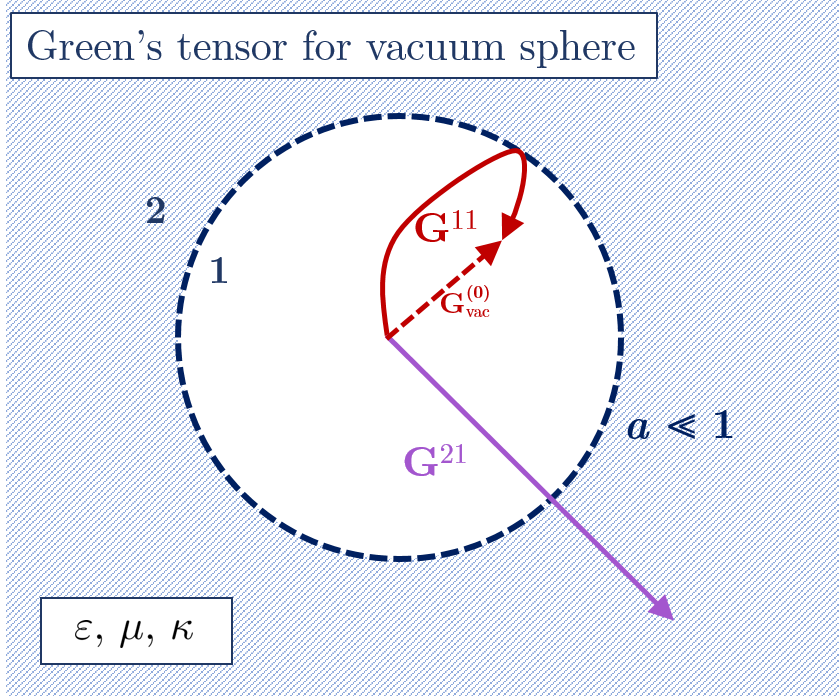}
\caption{Scheme for local-field correction calculation. The ansatz for a source point inside the sphere consists of $\G^{11}$ and $\G^{21}$, where the former corresponds to an absorption inside the sphere and the latter one to one outside. The Green's tensor $\G^{11}$ itself consists then of a bulk part $\Gfs$ and a scattering part.}
\label{lfc_scheme}
\end{figure}

We use the Onsager real cavity model to calculate the local-field corrections for spontaneous decay inside a chiral medium.
In this model, a free-space sphere is cut out around the interaction point, i.e.~the decaying atom at position $\bm r = 0$, inside the chiral medium. 
The Helmholtz-equation for the Green's tensor $\ten{G}(\bm r, \bm r', \omega)$ of such a system is then given by Eq.~\eqref{helmholtzsol} for $|\bm r|\ge a$ and by its free-space version ($\varepsilon = \mu = 1$, $\kappa =0$) for $|\bm r|< a$, where $a$ is the sphere radius. 
The boundary conditions of the electromagnetic field at $|\bm r| = a$ yield the boundary conditions for the Green's tensor
\begin{align}
\bm e_r \times
\ten{G}  (\bm r \rightarrow \bm a^- , \bm r' )
&= 
\bm e_r \times
\ten{G} ( \bm r \rightarrow \bm a^+ , \bm r' )
\,,
\label{ch:boundary1}
\\
 \bm e_r \times \nabla \times \ten{G} ( \bm r \rightarrow \bm a^-, \bm r' )
&= 
- \frac{\kappa \omega }{\mu }
\bm e_r \times
\ten{G} ( \bm r \rightarrow \bm a^+ , \bm r' ) 
\nonumber
\\
& \quad
+
\frac{1}{\mu } \bm e_r \times \nabla \times \ten{G} (\bm r \rightarrow \bm a^+  , \bm r' )
\,, 
\label{ch:boundary2}
\end{align}
for all $\bm a$ with length $a$ and where $\bm a^{\pm}$ denotes if the limit is carried out from left- or right-hand side.
\newpage
\begin{widetext}
We choose the ansatz for the Green's tensor
\begin{align}
r < a: \quad
\G^{11} (\bm r, \bm r' ) 
&=
\Gfs
+
\frac{\i }{4 \pi k_0 }
\sum_{\mathit{n}}^\infty \sum_{\mathit{m}}^{\mathit{n}}
(2 - \delta_{\mathit{m}0} ) \frac{(2n+1)(n-m)!}{n(n+1)(n+m)!}
\nn
&
\qquad\qquad\qquad 
\times 
\Big\lbrace
\big[ 
A_{\mathit{n}}^v \bm V_{\eo \mathit{mn} } (\bm r, k_0 )
+
A_{\mathit{n}}^w\bm  W_{\eo \mathit{mn} }  ( \bm r, k_0 ) 
\big]\bm  V_{\eo \mathit{mn} }(\bm r', k_0 )
\nn
&
\qquad\qquad\qquad\qquad 
+\big[ 
B_{\mathit{n}}^v \bm  V_{\eo \mathit{mn} }^{(1)} (\bm r,k_0 )
+
B_{\mathit{n}}^w \bm W_{\eo \mathit{mn} }^{(1)} ( \bm r, k_0 ) 
\big] \bm W_{\eo \mathit{mn} }(\bm r', k_0 )
\Big\rbrace
\,,
\label{ch:G11}
\\
r> a: \quad
\G^{21} (\bm r, \bm r' ) 
&=
\frac{\i }{4 \pi k_0 }
\sum_{\mathit{n}}^\infty \sum_{\mathit{m}}^{\mathit{n}}
(2 - \delta_{\mathit{m}0} ) \frac{(2n+1)(n-m)!}{n(n+1)(n+m)!}
\nn
&
\qquad\qquad\qquad 
\times 
\Big\lbrace
\big[ 
C_{\mathit{n}}^v \bm V_{\eo \mathit{mn} }^{(1)} (\bm r, k_+ )
+
C_{\mathit{n}}^w\bm  W_{\eo \mathit{mn} }^{(1)} ( \bm r, k_- ) 
\big]\bm  V_{\eo \mathit{mn} }(\bm r', k_0 )
\nn
&
\qquad\qquad\qquad\qquad 
+\big[ 
D_{\mathit{n}}^v \bm  V_{\eo \mathit{mn} }^{(1)} (\bm r, k_+ )
+
D_{\mathit{n}}^w \bm W_{\eo \mathit{mn} }^{(1)} ( \bm r, k_- ) 
\big] \bm W_{\eo \mathit{mn} }(\bm r', k_0 )
\Big\rbrace
\,,
\label{ch:generalSolution}
\end{align}
where $\Gfs$ is the free-space Green's tensor and the superscripts in $\G^{ij}$ denote the region $j$ of the source point $\bm r'$ and region $i$ of the absorption point $\bm r$ as labelled in Fig.~\ref{lfc_scheme}. 
\end{widetext}
The boundary conditions yield eight linear independent equations for the Green's tensor. Solving them, we obtain the coefficients $A_{\mathit{n}}^v$, $A_{\mathit{n}}^w$, $B_{\mathit{n}}^v$ and $B_{\mathit{n}}^w$ which vanish for all $n>0$ in the limit $a\rightarrow 0$.
\begin{widetext} 
The remaining ones are given by 
\begin{multline}
A_0^v
=
\frac{3 \i \left(\mu  (\mu +2)+(1-4 \mu ) n_{r}^2\right)}{2 a^3 k_0 (2 \mu +1) \left(\mu +2 n_{r}^2\right)}
\\
+
\frac{k_0^2 \left(-2 \mu ^2 (2 \mu +1)^2+36 \mu  n_{r}^7+9 \mu  \left(4 \mu ^2+8 \mu +1\right) n_{r}^5-8 (2 \mu +1)^2 n_{r}^4+9 \mu ^3 n_{r}^3-8 \mu  (2 \mu  n_{r}+n_{r})^2\right)}{2 (2 \mu +1)^2 \left(\mu +2 n_{r}^2\right)^2}
\\
-
\frac{9 \i k_0 \left(-\mu ^2 \left(5 \mu ^2+7 \mu +2\right)+20 \mu  n_{r}^6+\left(20 \mu ^3+32 \mu ^2-11 \mu -5\right) n_{r}^4-\mu  \left(11 \mu ^2+24 \mu +7\right) n_{r}^2\right)}{10 a (2 \mu +1)^2 \left(\mu +2 n_{r}^2\right)^2}
\\ 
+ \kappa \bigg(
\frac{18 k_0^2 \mu  \left((4 \mu +3) n_{r}^5+\mu  (3 \mu +2) n_{r}^3\right)}{(2 \mu +1)^2 \left(\mu +2 n_{r}^2\right)^2}-\frac{9 \i \mu }{a^3 k_0 (2 \mu +1) \left(\mu +2 n_{r}^2\right)}
\\
-\frac{9 \i k_0 \mu  \left(\mu  (3 \mu +1)+20 (\mu +1) n_{r}^4+\left(20 \mu ^2+23 \mu +3\right) n_{r}^2\right)}{5 a (2 \mu +1)^2 \left(\mu +2 n_{r}^2\right)^2}
\bigg)\,,
\end{multline}
\begin{multline}
A_0^w 
= 
\frac{9 \i \left(n_{r}^2-\mu ^2\right)}{2 a^3 k_0 (2 \mu +1) \left(\mu +2 n_{r}^2\right)}
-
\frac{9 \i k_0 \left(n_{r}^2-\mu ^2\right) \left(-\mu  (3 \mu +1)+20 \mu  n_{r}^4-(13 \mu +3) n_{r}^2\right)}{10 a (2 \mu +1)^2 \left(\mu +2 n_{r}^2\right)^2}
\\
+
\frac{9 k_0^2 \mu  \left(4 n_{r}^2-1\right) n_{r}^3 \left(n_{r}^2-\mu ^2\right)}{2 (2 \mu +1)^2 \left(\mu +2 n_{r}^2\right)^2} \,,
\end{multline}
 with $B_0^{v/w} =A_0^{w/v}\big|_{\kappa \rightarrow -\kappa}$.
\end{widetext}
With this we finally obtain the Green's tensor of interest,
\begin{align}
\nabla \times \Im \G^{\mathrm{lfc}} 
= \nabla \times \Im \G^{11}(\bm r = \bm r'=0 )
\,.
\end{align}
In smallest order of $\kappa$ and for small radius $a$ it is given by 
\begin{multline}
\nabla \times \Im \G^{\mathrm{lfc}} 
= 
\Im \bigg\{
\frac{ \kappa c }{ \pi a^3 \omega } f_3(\varepsilon,\mu )
+ 
\frac{   \kappa \omega }{ \pi a c}  f_1( \varepsilon,\mu)
\\
+
\i    \frac{\kappa \omega^2}{ \pi c^2}   f_0( \varepsilon,\mu)
\bigg\} \ten{I}
\,,
\label{curl_g_lfc}
\end{multline}
with $f_{\textit{i}}$ defined by 
\begin{align}
f_0 ( \varepsilon,\mu)
&= 
 \frac{3     \mu   \left(4 \mu  n_{r}^5+3 n_{r}^5+3 \mu ^2 n_{r}^3+2 \mu  n_{r}^3\right)}{ (2 \mu +1)^2 \left(\mu +2 n_{r}^2\right)^2} \;, \label{f0App}
\\
f_1( \varepsilon,\mu) 
&= 
\frac{ 3 \mu    }{10   (2 \mu +1)^2 \left(\mu +2 n_{r}^2\right)^2}
 \big[(\mu  (3 \mu +1)
 \nn
 & \qquad
 +20 (\mu +1) n_{r}^4+\left(20 \mu ^2+23 \mu +3\right) n_{r}^2 \big] \;,
\\
f_3( \varepsilon,\mu) 
&= 
\frac{3 \mu   }{2   (2 \mu +1)  (\mu +2 n_r^2 )}  \,.
\end{align}
The only contribution surviving in the imaginary part for real parameters is the last term proportional to $f_0$, which yields
\begin{align}
\nabla \times \Im \G^{\mathrm{lfc}} 
= 
\frac{3 \omega^2 \mu  \kappa\left((4 \mu +3) n_{r}^5+\mu  (3 \mu +2) n_{r}^3\right)}{\pi  c^2 (2 \mu +1)^2 \left(\mu +2 n_{r}^2\right)^2} \ten{I}
\,,
\end{align}
and is the same result as derived in Ref.~\cite{thesis_janine} where the local-field correction was derived for energy transfer. 
In case of complex-valued medium parameters, the result is dominated by the contribution proportional to $f_3$ (since $a \omega /c \ll 1$) and yields
\begin{align}
\nabla \times \Im \G^{\mathrm{lfc}} 
\approx \Im \left[ 
\frac{3 \kappa }{2 \pi  a^3 k_0 (2 \mu +1) (2 \varepsilon +1)}
\right]  \ten{I}
\,.
\end{align}
%
%


\section{Fresnel coefficients}\label{C}
\renewcommand{\theequation}{C.\arabic{equation}}
\setcounter{equation}{0}

We make use of reflection coefficients for achiral--chiral interfaces when the source is located at the achiral half space that were calculated in the Appendix A2 of Ref.~\cite{refcof}. In the notation used throughout the article for vacuum as the half space they are given by:
\begin{align}
r_{sp}=&-\frac{2\i}{D}\sqrt{\frac{\varepsilon}{\mu}}(a_+-a_-)\label{rsp} \, , \\[10pt]
r_{ps}=&-r_{sp}\label{rps} \, , \\[10pt]
r_{ss}=&\frac{(1+a_-)(1-\frac{\varepsilon}{\mu}a_+)+(1+a_+)(1-\frac{\varepsilon}{\mu}a_-)}{D} \, , \label{rss}\\[10pt]
r_{pp}=&\frac{(1-a_-)(1+\frac{\varepsilon}{\mu}a_+)+(1-a_+)(1+\frac{\varepsilon}{\mu}a_-)}{D} \, , \label{r_pp}
\end{align}
where $D$ is defined by
\begin{align}
D=&(1+a_-)\left(1+\frac{\varepsilon}{\mu}a_+\right)+(1+a_+)\left(1+\frac{\varepsilon}{\mu}a_-\right),\\
=&2+\left(1+\frac{\varepsilon}{\mu}\right)(a_-+a_+)+a_+^2+a_-^2 \;. 
\end{align}

The coefficients $a_-$ and $a_+$ for their part are given by
\begin{align}
a_-=&\sqrt{\frac{\mu}{\varepsilon}}\frac{k_0}{k^\perp}\frac{\sqrt{k_-^2-k^{\|2}}}{k_{-}} \, , \\
a_+=&\sqrt{\frac{\mu}{\varepsilon}}\frac{k_0}{k^\perp} \frac{\sqrt{k_+^2-k^{\|2}}}{k_{+}} \, ,
\end{align}
with $k_{\pm}$ as defined in Eqs.~(\ref{cwv}).


\bibliography{references}

\end{document}